\documentclass{aa}  
\usepackage{graphicx}
\usepackage{txfonts}

\def \nuwig {\nu_{\rm wgl}}
\def \twig {\tau_{\rm wgl}}
\def \nuobs {\nu_{\rm obs}}
\def \rhoc {\rho_{\rm crv}}
\def \phl {\phi_{\rm l}}
\def \pht {\phi_{\rm t}}
\def \til {t_{\rm l}}
\def \tit {t_{\rm t}}
\def \dtrot {\Delta t_{\rm rot}}
\def \dtobs {\Delta t_{\rm obs}}
\def \drrot {\Delta r_{\rm rot}}
\def \dr {\Delta r}
\def \dt {\Delta t}
\def \vel {\rm v}
\def \theb {\theta_{\rm R}}
\def \thh {\theta_{\rm h}}
\def \thm {\theta_{\rm m}}
\def \psimax{\psi_{\rm max}}
\def \dpsi{\Delta\psi}
\def \npsi{n_{\rm e}(\psi)}
\def \nel{n_{\rm e}}
\def \rlc {R_{\rm lc}}

\def \nuo {\nu_{\rm obs}}

\def \rhh  {\rho_{\rm h}}
\def \rh  {\rho_{\rm h}}

\def \sep {\Delta}
\def \imax {I_{\rm max}}
\def \imin {I_{\rm min}}
\def \dn {D_{\rm n}}
\def \dnl {D_{\rm n}^{\rm l}}
\def \dnt {D_{\rm n}^{\rm t}}
\def \cin {c_{\rm in}}
\def \cout {c_{\rm out}}

\def \rns{R_{\rm NS}}

\begin{document}

   \title{A model for double notches and bifurcated components
   in radio profiles of pulsars and magnetars}

   \subtitle{Evidence for the parallel acceleration maser in 
   pulsar magnetosphere}

   \author{J. Dyks
          \inst{1}
          \and
          B. Rudak\inst{1}
	  \and
	  Joanna M. Rankin\inst{2}
          }

   \offprints{J. Dyks}

   \institute{Nicolaus Copernicus Astronomical Center, Toru\'n, Poland\\
              \email{jinx@ncac.torun.pl}
             \email{bronek@ncac.torun.pl}
          \and
	    Physics Department, University of Vermont, Burlington, 
	    VT 05405, USA
             }

   \date{Received 30 October 2006 / Accepted 27 January 2007}

 
  \abstract
    {Averaged pulse profiles of three nearby pulsars:
    B1929$+$10, J0437$-$4715 and B0950$+$08 exhibit unusual `double notches'.
    These W-like looking features
    consist of two adjacent V-shaped dips
    that approach each other at increasing
    observation frequency $\nuobs$ roughly at a rate $\sep \propto \nuobs^{-1/2}$,
    where $\sep$ is the separation between the notches' minima.
   We show that basic properties of the notches, namely their
   W-like look and the rate of their converging
   can be understood within a narrow class of models of coherent
   radio emission from pulsars: the free electron maser models based on coherent
   inverse Compton scattering of parallel oscillations of ambient electric field.
   The observed properties of the pulsars imply that the Fourier spectrum
   of the wiggler-like oscillations is narrow and that the broad-band character
   of the radio emission reflects the width of the electron energy distribution.
   Such a model provides a natural explanation
   for the frequency-independent separation between the main pulse and interpulse
   of B0950$+$08 as well as for the lack of radius to frequency mapping
   in the conal-like emission of J0437$-$4715.
   The frequency behaviour of the
   main pulse in the profile of the first radio magnetar XTE J1810$-$197
   can also be explained within this model.

   }

   \maketitle


\section{Introduction}

Double notches are pairs of adjacent V-shaped dips 
observed in the averaged pulse profiles
of three nearby pulsars:
PSR B1929$+$10 (Rankin \& Rathnasree 1997), J0437$-$4715
(Navarro et al. 1997, hereafter NMSKB) and
B0950+08 (McLaughlin \& Rankin 2004, hereafter MR04).
The most striking property of the notches is that
they have the intriguing ``W" look:
both notches have similar (if not identical) width $W$
which is approximately equal to their separation $\sep$.
The separation decreases for increasing observation frequency $\nuobs$
(NMSKB).
The notches do not affect the degree of polarization nor its position angle.

Remarkably, the notches reside in weak and extended `pedestal' emission
that exhibits many unusual properties: 1) it is observed
far from the main pulse (MP), 2) covers long intervals of pulse longitude,
and 3) shows three strange polarization characteristics: 
3a) It is very strongly
linearly polarized ($\la$100\% for B1929$+$10). 3b) The
S-shaped position angle curve fitted to the pedestal emission only
(with the MP excluded) is shifted \emph{leftward} with
respect to the MP (eg.~Everett \& Weisberg 2001). 
This is the opposite shift direction than
expected for outward emission from a 
rotating magnetosphere (Blaskiewicz,
Cordes \& Wassermann 1991). 
3c) The behaviour of the position
angle is not very different from the predictions of the rotating vector
model (RVM, Radhakrishnan \& Cooke 1969), 
which is puzzling because of 1) and 2),
The properties 1) and 2) suggest that the pedestal radio
emission may originate from locations
in the magnetosphere that have little to do with the standard radio emission 
region. The closed field line region or 
extremely large altitudes, 
comparable to, or larger than, $R_{\rm lc}$ are not excluded.
In the case of J0437$-$4715 the notches are located in a wing of a seemingly 
conal component. However, the conal-like components in J0437$-$4715 
also exhibit
some special properties, eg.~they do not follow the radius to frequency 
mapping. Inferences of this paper refer to this special pedestal
and conal emission and should probably be not extended to
all known emission components, eg.~the core components.

Existing models of double notches interpret them as a double eclipse
of an extended emission region by a single absorber. The doubleness of 
such eclipse is caused by combined effects of differential 
(altitude-dependent) aberration and propagation time delays within
the spatially extended emission region (Wright 2004). The absorber/eclipter may 
corotate in outer parts of pulsar magnetosphere (Wright) or remain stationary
at the center of the magnetosphere (Dyks et al.~2005, hereafter DFSRZ).

The models based on the differential special relativistic effects
suffer from two main problems: 1) They provide no obvious reason
for the W shape of the notches. Our preliminary, simplified
calculations of pulse profiles for one version of such model (with the pulsar
as the eclipter, DFSRZ) failed to produce the W shape.
2) The large radial extent of the emission region in these models and the 
strong caustic effects associated with the mechanism
of the double eclipse should lead to rapid, complicated
variations of position angle and strong depolarization (Dyks et al.~2004a).
None of them is observed (in J0437$-$4715 they must be caused 
by interaction of orthogonal polarization modes.)
The simple property that $W \simeq \Delta$ is an extremely strong constraint
on any physical model of the notches, and our main goal was to devise one that
fulfills this requirement.
In Section \ref{secdata} we present the observed properties of double notches
of B1929$+$10 and compare it to the other pulsars. 
In Section \ref{secidea} we describe a general principle
of our model and the numerical code used to calculate modelled shapes
of notches. Section \ref{secdetails} describes how
the look of modelled notches depends on various model parameters.
Sec.~\ref{secmodcomp} furnishes our model with physics
and compares it with the observations.

\section{The double notches of PSR B1929$+$10}
\label{secdata}

Pulsar B1929$+$10 is a very useful
object for studying the double notches because: 
1) its notches are located far from the other strong 
emission components (MP, interpulse) and 2) the emission is strongly linearly 
polarized 
($\la100\%$ Rankin \& Rathnasree 1997).
Both properties ensure that the pedestal emission within
and around the notches is not contaminated by contributions from different
emission regions.
In this section we report new observations of B1929$+$10 performed by one of us
(JR) at Arecibo Observatory at frequencies $327$ MHz, $1.17$ GHz, 
and $1.5$ GHz with the respective bandwidths of $25$, $100$ and $200$ MHz.

   \begin{figure}
      \centering
      \includegraphics[width=0.5\textwidth]{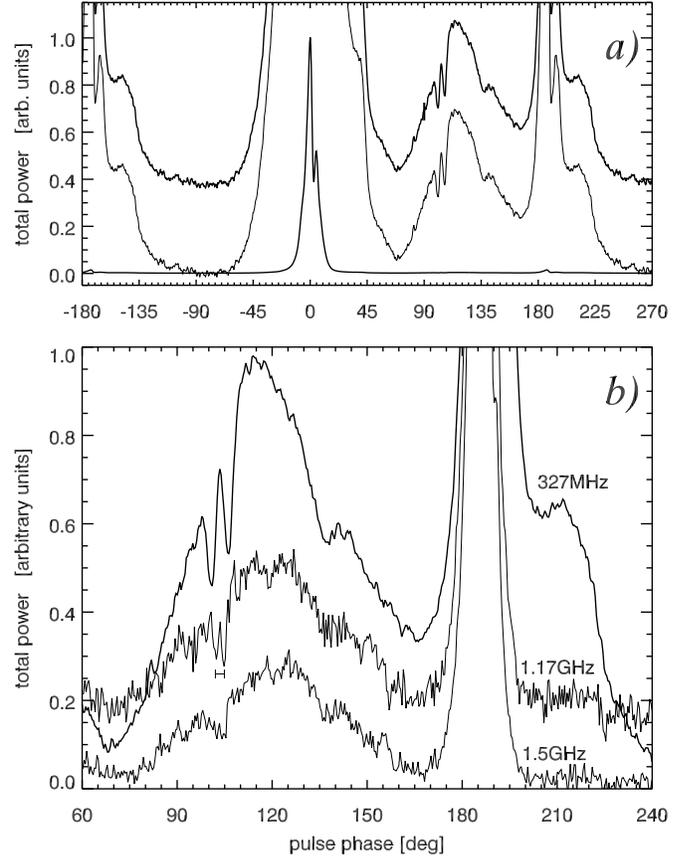}
      \caption{Averaged pulse profiles of B1929$+$10 (all lines present
      the total flux). 
      {\bf a)} Three representations of the $327$-MHz profile
      that show how the baseline level affects the depth $\dn$ of the notches.
      The top line has the minimum recorded flux $\imin$ 
      set at $0.0015\imax$ which results in $\dn\simeq 20\%$.
      The notches are much deeper ($\dn\simeq 37\%$) if the profile 
      is shown in normal way (thin middle line with $\imin=0$) . 
      For both these profiles $\imax=250$. 
      The profile at the bottom has $\imax=1$ and $\imin=0.0015\imax$.
      {\bf b)} Double notches at three different frequencies $\nuobs$.
      The horizontal bar below the notches at $1.17$ GHz marks the separation
      predicted by eq.~(\ref{sepnu}) for $\sep_{\rm 327 MHz}=5.36^\circ$.
      The phase alignment is such that the MP maxima at $1.17$ and $1.5$ GHz
      lag the MP maximum at $327$ MHz by $0.8^\circ$. Different vertical scales
      and baseline levels were used at different $\nuobs$ for viewing purposes.
      }
      \label{datafig}
   \end{figure}
										    
   \begin{figure}
      \centering
      \includegraphics[width=0.5\textwidth]{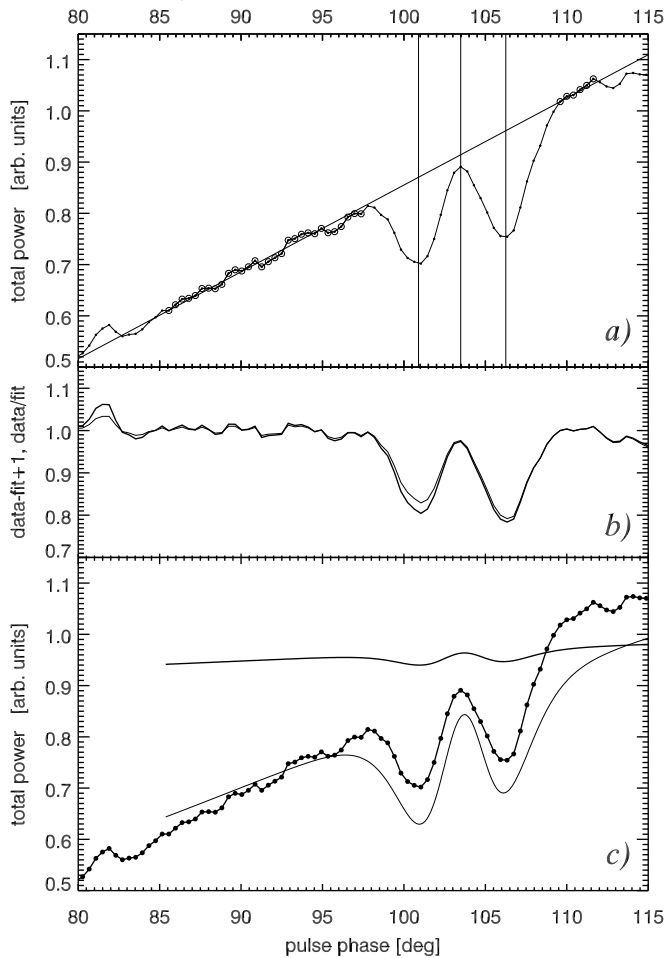}
      \caption{A zoom into the $327$ MHz notches of B1929$+$10
      (expanded part of the top line in Fig.~\ref{datafig}a).
      The circled points in {\bf a)} have been used to fit the linear
      variations of flux around the notches. We measure the depth of notches 
      $\dn$ along the outer vertical lines; the separation $\sep$
      is the horizontal distance between them. 
      {\bf b)} The notches with the linear trend removed. The thin line
      shows $\rm data - \rm fit + 1$; the thick line presents the ratio 
      $\rm data/fit$. The leading notch is slightly less deep than 
      the trailing one. {\bf c)} One of model profiles (thick solid) 
      obtained for the $\vec a \parallel \vec \vel$ case with $\gamma=10$
      plotted over the data. The modelled notches are some $\sim 10$ times
      shallower than the observed ones. The thin solid line presents
      the same model result after a linear rescaling of the y-axis,
      which makes the modelled notches quite similar to the data.
      }
      \label{modcomp}
   \end{figure}

Fig.~\ref{datafig}a presents the averaged pulse profile of 
B1929$+$10  at $327$ MHz.
The double notches are visible
at $\phi=103.5^\circ$.
It will become clear below that 
a particularly constraining parameter for
theoretical models is the
depth $\dn$ of the notches. Unfortunately, this quantity 
is also the most difficult to determine from observations 
because it depends on the amount of flux
received from the pulsar at the lowest (dimmest) point of its pulse profile.
This `unpulsed emission' was detected in 
B1929$+$10 via phase-resolved interferometric 
observations at 408 MHz (Perry \& Lyne 1985, hereafter PL85).

To account for the depth measurement problem Fig.~\ref{datafig}a 
represents the profile in a few different ways:
The middle line (thin) is represented
in the usual way, with the zero of the y-axis at the level 
$\imin$ of the
lowest place in the pulse profile, which we take 
as an average of data within 
the phase range $(-90,-80)$. For the top line 
we assumed that $\imin = (1.5\cdot 10^{-3} \pm 1.5\cdot 10^{-4})\ \imax$,
ie.~we take the same ratio of $\imin/\imax$ at 327 MHz 
as observed by PL85 at 408 MHz. The error bar at $\phi=90^\circ$
marks the 2$\sigma$ error of the baseline level from PL85.
The actual error may be larger if $\imin/\imax$ is strongly frequency 
dependent.
The top two lines are multiplied by a normalization 
factor that sets the maximum observed
flux $\imax$ at 250 (beyond the figure) 
to reveal the shape of the weak pedestal emission.
For the third, bottom pulse profile 
$\imax = 1$ and $\imin/\imax=1.5\cdot 10^{-3}$.
The upper two curves illustrate how strongly the derived depth 
of the notches depends on the baseline level: the tiny shift of the baseline
from zero up to
$1.5\cdot 10^{-3}\imax$ decreases $\dn$ from nearly $40\%$ (middle curve)
down to $20\%$ (top).  
Interestingly, the flux of the pedestal emission
increases roughly linearly with time (angle) near the notches.

The 327-MHz pulse profile consists of
two observations (12238 and 18835 single pulses long).
A comparison of averaged pulse profiles for these two
pulse sequences suggests that
the relative intensity of the various weak structures in the pedestal
may vary in time. 
The values of $\dn$ derived for these two observations are equal to 
$17\%$ and $24\%$ (assuming $\imin/\imax$ from PL85).
It is hard to tell whether this is caused by scintillation
or intrinsic variability.

Fig.~\ref{datafig}b shows the frequency evolution of a part 
of the profile.
The notches approach each other at $1.17$ GHz (middle curve)
and seem to be merged into a single feature at $1.5$ GHz (bottom).
Note that each of the profiles in Fig.~\ref{datafig}b has undergone
different linear transformation to fit a single plotting box
(no meaning should be attached to the depth of the notches).
The profiles were phase-aligned in  
such a way that the MP maximum of the 327-MHz profile
in our Fig.~\ref{datafig}b
precedes in phase the MP maximum of the 1.4-GHz profile by 
$\Delta\phi_{\rm \tiny MP} \simeq 0.8^\circ$.
Similar misalignment of the MP maxima can be discerned 
in published figures that present 
time-aligned profiles at different frequencies.
From fig.~2 of Kuzmin et al.~(1998), that was obtained for $\rm DM = 3.18$
pc cm$^{-3}$,
one can estimate $\Delta\phi_{\rm \tiny MP} \simeq 1.3^\circ$.
In fig.~12 of Hankins \& Rankin (2006) 
$\Delta\phi_{\rm \tiny MP} \simeq 0.3^\circ$
for $\rm DM = 3.176$ pc cm$^{-3}$.
Interestingly, the shift of $0.8^\circ$ puts into perfect alignment 
\emph{both} the notches and the maximum of the interpulse. 
Given that the shape of the MP changes strongly between 0.4 and 1.4 GHz,
it is reasonable to assume that the phase location of the MP maximum
in B1929$+$10 is frequency dependent whereas the location of notches
is fixed.
Position of double notches definitely 
does not depend on frequency in J0437$-$4715 (NMSKB) nor in B0950$+$08 (MR04).
All existing observations of notches
are then consistent with their location in pulse profiles being 
frequency-independent. A physical explanation for this will be proposed 
in Section \ref{secmodcomp}.

Fig.~\ref{modcomp} presents a zoom into the same 327-MHz notches
as shown in Fig.~\ref{datafig}a (top curve). Note that their appearance 
(especially the sharpness) is very sensitive to the assumed range
on both the vertical and horizontal axis. Panel \ref{modcomp}a presents
a linear fit to the marked data points around the notches (circles).
The depths $\dn$ are measured between this fit
and the notches' minima as marked with the
outer vertical lines. The result is
$\dnl = (20.2\pm1)\%$ for the leading notch and $\dnt=(21.8\pm1)\%$ for the trailing one
(when expressed in percent, the depth of a given notch is normalized
by the value of the linear fit at the phase at which the
notch has the minimum; 2$\sigma$ errors are used throughout this paper).
In Fig.~\ref{modcomp}b the linear fit is subtracted from the
data in two ways: the thin line represents $\rm data - fit + 1$,
whereas the thick line shows the ratio $\rm data/fit$.
In either case, the leading notch (located in the weaker
emission) is shallower than the trailing notch in the
stronger emission. The asymmetry holds at higher $\nuobs$. 
With $\phi=0$ set at the MP maximum, the notches are located at 
$\phi_{\rm l}=100.9\pm0.07^\circ$ and $\phi_{\rm t} = 106.25\pm0.06^\circ$,
ie.~they are separated by $5.36\pm0.1^\circ$. The center of W lags the MP
by $103.5\pm0.1^\circ$.

An important feature that strongly constrains the theoretical models
is that the maximum between the double notches (at the center of the `W')
nearly (but not exactly) reaches the level interpolated with the linear fit.
The data point marked in Fig.~\ref{modcomp}a with the central vertical line
($\phi=103.5^\circ$) is $2.6\%$ below the fit level, ie.~it misses the fit
by $\sim 10\%$ of $\dn$.

\subsection{Notches in other pulsars}

In the millisecond pulsar J0437$-$4715 the emission
with notches is stronger ($\sim 8\%$ of $\imax$ at $438$ MHz)
and a bit less polarized ($\sim 70\%$) than in the case of B1929$+$10.
The notches actually seem to be carved out 
of the trailing wing of a prominent trailing component in the main pulse
(see figures in NMSKB).
At sub-GHz frequencies the top of the trailing component 
is split
into two maxima separated by a single dip/notch. 
This bifurcated top is probably bright enough (at least at $\nuobs \la 400$ MHz)
to be studied on a single pulse basis (see Sec.~\ref{secsingle}).
We are not aware of a baseline flux measurement for this object,
so that only the upper limit of $\sim50\%$ for the notch depth can be 
estimated. 
If the unpulsed flux from this pulsar is comparable to a few percent 
of $\imax$ then the actual $\dn$ is considerably smaller.

As in the case of B1929$+$10, the double notches in J0437$-$4715 
approach each other for increasing frequency $\nuobs$, 
though they are still separable
in the high quality $1.5$-GHz pulse profile of NMSKB.
The split at the top of the trailing component also gets narrower
at increasing frequency and looks nearly like a single feature at 
$1.5$ GHz

On the theoretical side, the pulse profile of J0437$-$4715 may be more
difficult to model because its magnetosphere is much smaller
($P=5.76$ ms, $\rlc=27.5\cdot 10^6$ cm) 
than that of B1929$+$10 ($P=0.2265$ s, $\rlc=1080\cdot 10^6$ cm)
so that the radio emission has a larger 
probability of being affected by general relativistic effects.

Observations of B0950$+$08 performed by MR04
at 430 MHz give $\dn \sim 10 - 16\%$ and PL85 report negligible 
amounts of unpulsed emission for this object at a similar frequency.
However, the notches of B0950$+$08 are located close to the MP, 
in a region
probably contaminated by several types of emission, as the low
polarization degree suggests. It is therefore not excluded 
that the actual $\dn$ is larger than given above. 
On the other hand, the weak emission components
in B0950$+$08 have been reported to vary on a time scale of 
several days (MR04), and it is hard to tell at what stage of this variability
the observations of PL85 were done. The notches of B0950$+$08
seem fairly blurred at high frequencies (see inset in fig.~3 in MR04).

   \begin{figure*}
   \resizebox{12.7cm}{!}{\includegraphics[width=0.75\textwidth]{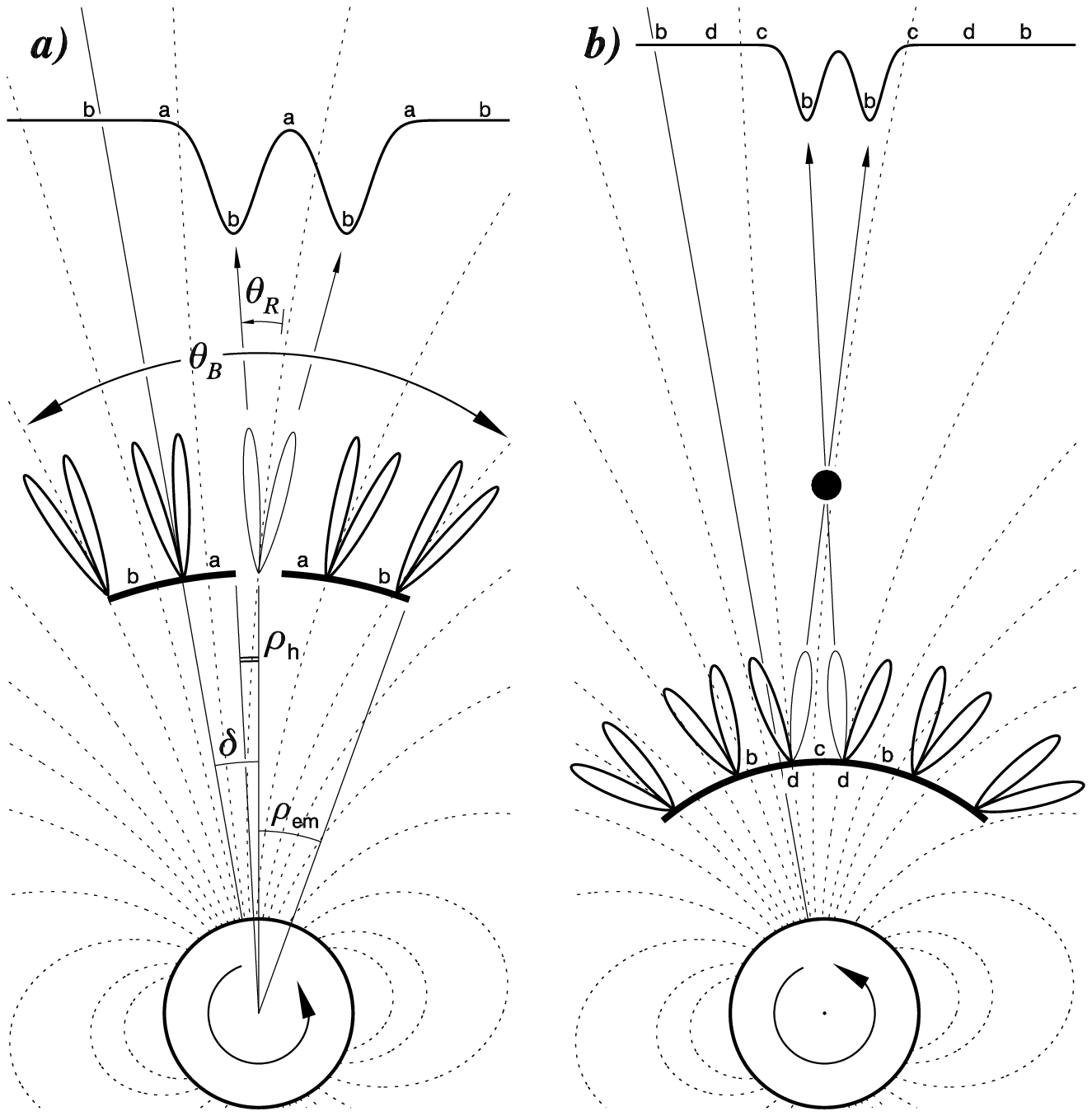}}
   \hfill
   \parbox[b]{50mm}{
      \caption{The principle of our model.
      The plane of rotational equator for a pulsar with dipole inclination
      $\alpha=90^\circ$ chosen for simplicity is shown.
      The thick solid arc is the crossection through
      a two-dimensional emission region.
      Crossections through several hollow cone elementary beams of radio 
      emission
      are indicated. The observer is located near the plane
      of the figure. The dotted curves mark the dipolar magnetic field lines.
      In {\bf a} the notches form because of the hole in the emission region.
      In {\bf b} they are created by a single absorber
      above the region (marked with bullet).
      The small letters help to associate different points of emission region
      with pulse phase at which the radiation emitted from them is observed.}
         \label{idea}
	 }
   \end{figure*}

\section{The model}
\label{secidea}

The model postulates an emission region with large extent
in the rotational azimuth $\phi$ and rotational colatitude $\theta$
($\Delta\phi \gg \Delta$ and $\Delta\theta \gg \Delta$).
For simplicity, we assume the emission is produced at a fixed altitude
(see Sect.~\ref{secemigeom} for more comments on emission region).
The key assumption is that
on a microscopic scale
the region radiates a hollow cone beam along a local
direction of magnetic field $\vec B$.
A single hole in the emission region produces a cone of reduced
emission, as shown in Fig.~\ref{idea}a. The notches can be observed
when the observer's line of sight cuts
through the cone.
A single absorber between an observer and the source of radio emission
will act in a similar way (Fig.~\ref{idea}b). The part of the emission
region that is ``half-hidden" from the observer has the shape of a ring.
Its crossection is marked with `d' in Fig.~\ref{idea}b.

There are at least two basic
reasons for which the elementary emission beam
can have the hollow cone shape:\\ 
1) Electrons may be accelerated
along their velocity $\vec \vel \parallel \vec a$
and the radio emission observed in the pedestal
can be due to this acceleration
(Melrose 1978; Kunzl et al.~1998; Schopper et al.~2002; Levinson et
al.~2005). 
In this case the opening angle of the elementary beam $2\theb \sim
1/\gamma$, where $\gamma \sim 10$ is the Lorentz factor of the electrons.
More exactly the received power has a maximum at the angle
\begin{equation}
\theb = \arccos \left[\frac{(1 + 24\beta^2)^{1/2} - 1}{4\beta} \right]
\approx \sqrt{\frac{1}{5}} \cdot \frac{1}{\gamma} \approx 0.4472\cdot 
\frac{1}{\gamma},
\label{textbook}
\end{equation}
where $\beta=\vel/c$ is the electron speed in units of the speed of light
and the approximation holds for $\gamma \gg 1$.
In this version of the model the observed separation
of the double notches is:
\begin{equation}
\sep \approx \frac{2\theb}{\sin\zeta} = \frac{2}{\sqrt{5}\ \gamma\sin\zeta}
= \frac{0.8944}{\gamma\sin\zeta} \approx \frac{1}{\gamma}
\label{separation}
\end{equation}
where $\zeta$ is the viewing angle measured from the rotation axis,
the factor $1/\sin\zeta$ takes into account
the `not a great circle' effect (eg.~Dyks, Rudak \& Harding 2004b),
and the latter approximation holds with accuracy better than $10\%$
for any angle $52^\circ < \zeta < 128^\circ$.
The equation is approximate in that it does not take into account
the angle $b$ between the hole axis and the line of sight. Numerical results 
of the next section tell us that this `hole impact angle'
does not affect $\sep$ as long as the notches have the W-like shape
(see Fig.~\ref{parallel}d). Eq.~(\ref{separation}) is valid for the
case of the emission region with a hole (Fig.~\ref{idea}a).
Note that according to eq.~(\ref{separation}) 
the observed magnitude of $\sep \sim 2^\circ
-6^\circ$ implies Lorentz factors of the order of 10,
which is a value estimated on independent grounds in pulsar
models based on the parallel acceleration
(eg.~the ALAE model of Melrose 1978 or
CICS model of Schopper et al.~2002). 

2) The elementary beam has the hollow cone shape also when 
high-energy electrons ($\gamma \gg 10$)
are accelerated perpendicularly ($\vec a \perp \vec \vel$)
but most of them have non-negligible pitch angle $\psi \gg 1/\gamma$
(see eg.~fig.~6.5 in Rybicki \& Lightman 1979).
In this case $\theb = \psimax$,
where $\psimax$ is the pitch angle for which the distribution of electrons
$n_{\rm e}(\psi)$ has a maximum. The non-zero pitch angle requires low
magnetic field, such as is present at large altitudes, comparable to
$\rlc$ (Lyubarski \& Petrova 1998; Malov \& Machabeli 2001; Petrova 2003;
Harding et al.~2005) as well as at lower altitudes in the millisecond
pulsars (eg.~J0437$-$4715). In spite of some positive features
(see.~Appendix \ref{pitchang}),
the pitch angle version of our model is less successful in reproducing
the data and is disfavored in this paper.

\subsection{The numerical code}
\label{numcode}

The numerical code to simulate the notches assumes:
1) specific geometry and size of the emission region;
2) specific geometry and size of the hole/fissure or absorber;
3) the structure of the magnetic field within the emission region, eg.
dipolar, radial (swept out by wind) or toroidal (sweepback);
4) the geometry of the elementary emission beam;
in the case $\vec \vel \parallel \vec a$ it
is calculated directly from the classical electrodynamics formulae
(eg.~eq.~4.101 in Rybicki \& Lightman 1979)
for a selected value of $\gamma$ (we do not integrate
over the electron-energy distribution).
In the case of $\vec \vel \perp \vec a$ we assume the shape of the
pitch angle
distribution (eg.~$n_{\rm e} \propto \psi\ \exp[-\psi^2/{\psimax}^2]$,
as in Epstein 1973),
and integrate the single electron `pencil' beams of radiation
over the pitch angle
distribution to get the ``elementary" emission beam
for a fixed value of $\gamma$.
Next, the emission region is divided into a large number of tiny fragments,
each of which is equipped with its own elementary beam. An observer
located at a viewing angle $\zeta$ is selected and a lightcurve
is calculated by integration over the source for each pulse longitude
(the place where the observer's line of sight enters the elementary beam
is determined individually for each fragment of the source).
The symmetry axis of the elementary beams is assumed to point along
the local magnetic field $\vec B$ in the corotating frame.
The aberration and propagation time delays are neglected
because of the fixed emission altitude.

To integrate the received flux over the electron-energy distribution
one needs to know how the coherent emissivity $\epsilon$
at a specific frequency $\nuobs$
depends on the electron Lorentz factor $\gamma$. This depends 
on the radio emission/coherency mechanism that actually works
in pulsar magnetosphere. Even for a specific emission mechanism,
the dependence of $\epsilon(\gamma)$ at a fixed $\nuobs$ may be completely
different for different values of $\nuobs$. 
Let us take the noncoherent curvature radiation
(that certainly \emph{cannot} be responsible for the emission with notches) 
as an example. Well below the maximum in the curvature
spectrum ($\nu \ll \nu_{\rm CR}$) the emissivity does not depend on $\gamma$,
whereas above the maximum
(in the region of the exponential cut-off) it becomes extremely
sensitive to it: $\epsilon_{\rm CR} \propto 
\exp[-const\cdot\gamma^{-3}]\cdot\gamma^{-1/2}$.
For a weak dependence of $\epsilon(\gamma)$ 
the integration over the electron energy distribution
could significantly blur the notches.
It is therefore very important that in the model of the parallel 
acceleration maser 
based on the coherent scattering of the `wiggler-like' field,
the emissivity is tightly focused around 
$\nuobs \sim \gamma^2\nuwig$
(see left panel of fig.~4 in Schopper et al.~2002),
where $\nuwig$ is the frequency of oscillations of the ambient
electric field.

The angular distance $\thh$ of the hole from the rotation axis 
and the dipole inclination $\alpha$ 
have little effect on the shape of notches, except from blowing
them up by the factor $1/\sin\thh$. Therefore,
in all cases presented in this paper we assume 
that $\thh=\alpha=90^\circ$. The emission region 
in all simulations of notches is a fragment of a sphere
that extends significantly both in $\phi$ and $\theta$ direction
($\Delta\phi = \Delta\theta = 10\theb$).

\section{Numerical results}
\label{secdetails}

We have modelled numerous configurations with different hole
geometries, absorber, $\vec B$-field structure and with
various pitch angle distributions $n_{\rm e}(\psi)$.
In general, results are sensitive to bulk geometry/topology of the
$\vec B$-field and emission/absorption region. In the pitch angle
case they also depend on the form of function $n_{\rm e}(\psi)$.

\subsection{The parallel acceleration case -- a hole in the 
dipolar $\vec B$-field}

\begin{figure*}
   \resizebox{12.7cm}{!}{\includegraphics[width=0.8\textwidth]{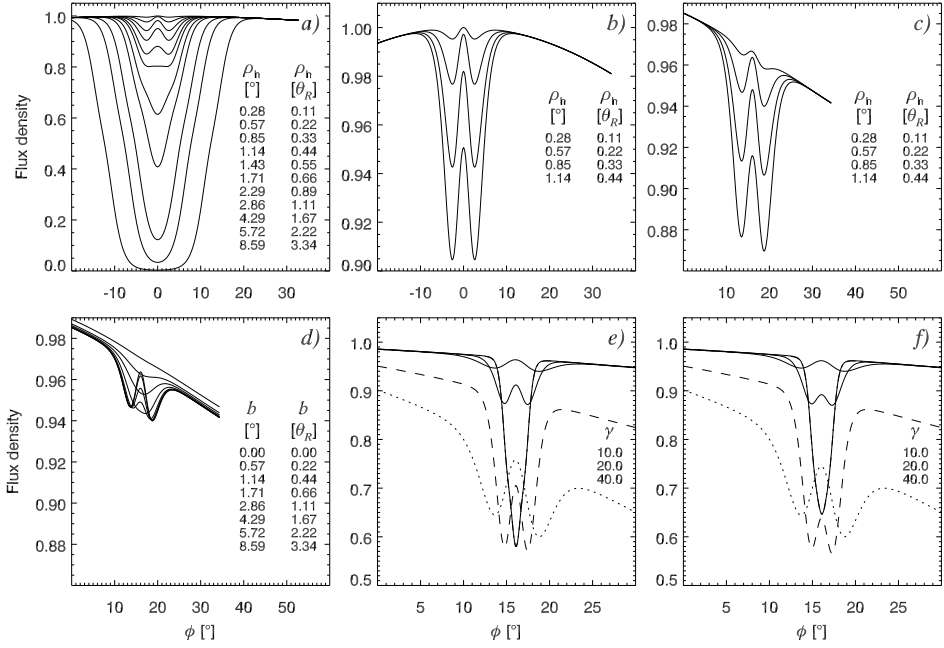}}
   \hfill
   \parbox[b]{50mm}{
      \caption{Notches modelled for a hole in the dipolar $\vec B$-field
      for the case $\vec a \parallel \vec \vel$.
      {\bf a)} The effect of increasing hole radius
      $\rho_{\rm h}$ for $\gamma=10$, $b=90^\circ-\zeta=0$ and $\delta=0$,
      ie.~with the hole centered at the dipole axis (top to bottom).
      {\bf b)} Top four curves
      from a) replotted with a rescaled vertical axis. {\bf c)} Same as b)
      but with the hole located off-axis ($\delta=30^\circ$).
      {\bf d)} The effect of increasing `impact' angle $b$
      for $\rho_{\rm h}=0.22\theb$ and $\gamma=10$. {\bf e)} The effect of
      increasing
      $\gamma$ for $\rho_{\rm h}= 0.57^\circ$ and $b=0$ (solid curves). 
      The lines for $\gamma=
      10$ and $20$ were linearly transformed and replotted as
      the dotted and dashed curves, respectively. {\bf f)} Same as e) but for
      $b=0.57^\circ$. Unlike in many figures that show \emph{observed} 
      pulse profiles, in all
      cases shown with solid lines the zero of y-axes corresponds to `no flux'.
              }
         \label{parallel}
	 }
   \end{figure*}

Fig.~\ref{parallel} presents results obtained for 
a circular hole in the dipolar magnetic field (the case
$\vec \vel \parallel \vec a$ with $\gamma=10$).
In \ref{parallel}a the hole is centered at the dipole axis
($\delta=0$),
and its angular radius $\rh$ increases from $0.1$ to $3.3\theb$
(top to bottom). The observer was located at a `hole impact angle'
$b \equiv \thh - \zeta = 90^\circ - \zeta = 0$, 
ie.~the line of sight sweeps through
the hole's center.
For $\rhh < 0.3\theb$ the notches have the  `W'-like shape similar 
to the observed one, with the flux at the center of the `W' at nearly
the same level as beyond the notches. 
An important difference between these cases and observations is that
the depth of the modeled notches does not exceed a few percent,
in comparison with a few tens of percent tentatively derived from
observations. This is a major concern for the present version of the model
and will be discussed below.
For a larger $\rh$ the central
flux drops down until a single wide ``notch" with an initially flat
bottom appears at $\rh \sim 0.7\theb$. 

In \ref{parallel}b the top four curves from \ref{parallel}a are replotted
with a stretched y-axis. This shows how sensitive the appearance
of the notches is to the selected viewing method. The strong stretch
of the y-axis produces the fall-off of 
flux for the increasing $|\phi|$. This occurs because the dipolar magnetic 
field lines are more spreaded at larger magnetic colatitudes
$\thm$ measured from the dipole axis (so that a smaller number of
lines are pointing towards a unit solid angle at increasing $\thm$
whereas the emissivity of the emission region is
assumed to be uniform per unit surface).
Note that a similar stretch of the y-axis may be unknowlingly
applied to plots that show the observed pulse profiles if the
baseline level has been overestimated (Fig.~\ref{datafig}a).

Fig.~\ref{parallel}c shows the effect of a non-axisymmetric
location of the hole: the dipole axis has been rotated by the angle 
$\delta=30^\circ$  with respect to the line connecting
the centers of the hole and star (see Fig.~\ref{idea}a).
According to the basics of dipolar field geometry, the notches
moved to $\phi \approx 0.5\delta$. 
The displacement of the hole off the dipole axis results
in moderate asymmetry of the notches' shape. The approximately
linear decrease of flux around
the notches is caused by the increasing spread of $\vec B$-field lines. 

Fig.~\ref{parallel}d presents the shape of notches for different
viewing angles $\zeta=\thh-b$ and the other parameters ($\rhh$, $\gamma$)
the same as in the second-from-top case in panel c. 
For $b < 0.3\theb$ the `W'-like
notches can be observed. For larger $b$ their shape evolves into a single dip.
Interestingly, 
the separation between the notches practically does not depend
on $b$ as long as $b < 0.7\theb$.
The separation starts to decrease only for $b \la \theb$, ie.~when the
flux at the W center becomes nearly as low as at the notches' minima. 
This property allows us to ignore the 
ratio of $b/\theb$ 
in eq.~\ref{separation} as well as in the discussion of 
the frequency dependence of $\sep$ in Section \ref{secmodcomp} below
(note that $b/\theb$ can change with $\nuobs$ for a single object
viewed by a fixed observer).

For the increasing $\gamma = 10$, $20$, $40$
the modelled notches approach each other, the flux at the W center
decreases, and they finally merge into a single feature
(solid lines in panels e and f of Fig.~\ref{parallel}).
This resembles the behaviour observed at different frequencies
(Fig.~\ref{datafig}b). It can be directly interpreted within our model
if radiation at higher frequencies
is mainly generated by electrons with larger Lorentz factor $\gamma$. 
The depth of the modelled feature increases when the notches merge.
This does not seem to happen in the data (Fig.~\ref{datafig}b)
although the baseline level at the $1.17$ and $1.5$ GHz is not known.
As we show in the next section, however, the interpretation is
supported if our geometric idea is supplied with the physics of
parallel acceleration maser.

As we show in Appendix \ref{pitchang} one can increase
the depth of notches up to $\sim 20$\% in the pitch angle scenario
by varying the function $n_{\rm e}(\psi)$. 
However, this is usually associated
with considerable deformation of their shape.
Modelled notches with a depth and shape reasonably similar to the observed one
can be obtained only after application of artificial ``linear rescaling"
of model lightcurves. 
Fig.~\ref{modcomp}c shows the result of such transformation
for the parallel acceleration case. 
The thick line with 
dots shows the observed $327$-MHz notches of B1929$+$10 
with the baseline level corrected according to the $410$-MHz measurement
of PL85. The thick solid line with the very shallow
notches is the second-from-top result from 
Fig.~\ref{parallel}c after a mirror reflection and horizontal shift
has been applied to it (with no rescaling in any axis).
The thin solid line in Fig.~\ref{modcomp}c is the same model result
after a linear rescaling of $y$ axis. This exercise is to check whether
the shape of the rescaled notches in the $\vec a \parallel \vec \vel$ case
bears any resemblance to the observed ones. 
The agreement is moderate: the outer sides of the modelled notches
are less steep than the observed one.

\subsection{Geometry of the hole/absorber}

\begin{figure}
      \centering
      \includegraphics[width=0.5\textwidth]{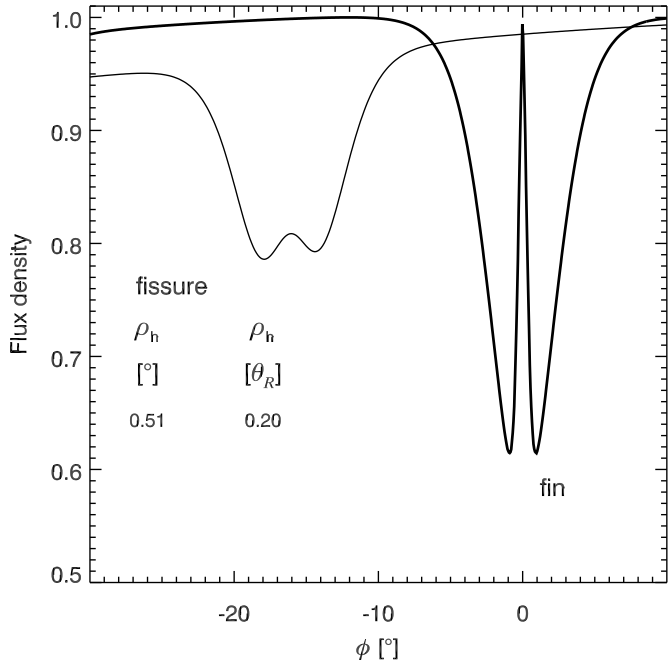}
      \caption{Modelled double notches for a meridionally extended
       fissure (thin line) and for a meridionally extended fin that protrudes
       vertically upward from the emission region (thick line). 
       The notches have
       large depth but their shapes do not resemble the observed one. 
	}
      \label{config}
   \end{figure}

The geometry of the hole or absorber is the next important factor
that affects the shape and depth of the notches.
We have considered a few cases that produce deeper notches.
However, this has always been accompanied by the deformation
of their shape.
An obvious way to make the notches deeper is to extend the
hole in the direction perpendicular to the rotational azimuth.
This, however, strongly decreases the flux at the W center
(Fig.~\ref{config}, thin line).
Much the same result is obtained when an elongated absorber 
(dense plasma stream?) is placed above the emission region.
A desperate way to avoid the decreased central flux
is to assume an opaque \emph{thin} wall/fin protruding
from the emission region upward. Such configuration
can produce very deep notches with the central flux unaffected
(if the fin is thin). 
However, the notches have a shape different than observed ($\sep \ll W$)
and become extremely asymmetric if the fin is located asymmetrically
in the dipolar magnetic field (off the dipole axis). Though the fin cannot
be responsible for the observed notches (and it would be hard to 
justify its origin) this case shows that a geometric configuration
that gives deep notches with the unaffected central flux is possible.

\subsection{Magnetic field geometry}

In addition to the dipolar $\vec B$ we have considered 
the radial and toroidal magnetic field. 
The radial field
has similar geometry to the dipolar field near the dipole axis
and it produced similar results.
The toroidal field ($B_x = \cos\phi, B_y = \sin\phi, B_z=0$)
was considered as the idealized model of the swept-back, near-$\rlc$
magnetic field. The direction of such $\vec B$ at a fixed azimuth
is independent of $\theta$ so that all parts of the extended source 
that are located at the meridian selected by the line of sight 
contribute equally to the observed flux. This tends to
decrease $\dn$.

\subsection{Emitter's geometry}
\label{secemigeom}

Our present choice of the emitter was dictated
by two facts: 1) the weak pedestal emission that contains the notches
covers a very large range of pulse phase. This implies that the angular
extent of the emitter must be large in $\phi$, \emph{and consequently in} 
$\theta$.
A negligible extent in $\theta$ would make the detection 
of the emission less probable, whereas the extended emission components
seem to be quite common among the nearest and brightest pulsars.
From the inspection of profiles in the EPN pulsar data base 
one can learn that within the distance of the furthest pulsar with notches
(B1929$+$10 at $0.36$ kpc) only $50\%$ of pulsars in the ATNF catalogue
(Manchester et al.~2005) has the radio emission limited to a narrow 
range of pulse phase (few tens of degrees), 
and some objects have really wild pulse profiles
(eg.~J2124$-$3358, Manchester \& Han 2004). 
2) The PA curve for the pedestal emission closely follows
the curve of Komesaroff (1970) which probably implies that the source
is not very extended in the radial direction ($\Delta r \ll \rlc$).
The fixed emission altitude that we have assumed
can be considered a good approximation for any region
with the radial extent that is negligible
in comparison with other relevant length scales.
These include the scale $\rhoc\Delta$ on which the magnetic field changes direction 
by angle $\Delta$, where $\rhoc$ is the radius of curvature of magnetic field lines.
The other scale is the one on which the electron plasma frequency drops significantly
(for a power-law decrease of plasma density it is of the order 
of the emission height $r$). Both these length scales are comparable.
Strong, fixed frequency radio emission from a `too large' range 
of altitudes would probably
make the notches less distinct.
The only emitter that we have managed to consider so far 
(ie.~the part of spherical surface shown in Fig.~\ref{idea}) surely does not
exhaust all the possible configurations that can produce
the extended pedestal and have a reasonable detection probability.
While simulations for other emitters are being considered,
we turn to the observational consequences
of physical aspects of the model. 

\section{Model versus observations}
\label{secmodcomp}

Based on a large number of numerical results, we find that
a few key features of model notches agree well
with the observations:
\begin{enumerate}
\item Whenever the sightline cuts through the cone of reduced
emission the calculated profiles tend to have two minima/notches;

\item Both notches have similar depth;

\item The width of the notches tends to be equal to their separation. 
\end{enumerate}

\noindent The list can be considerably
extended if our simple idea is supplemented with the physics
of a specific version of a free electron maser:

\begin{enumerate}
\item[4.]  If the emission at a higher frequency $\nuo$ is mainly
generated by electrons with larger Lorentz factors,
then the elementary emission beam is narrower at larger $\nuo$
which directly implies that the separation of notches is smaller,
as is indeed observed.
In the case of $\vec a \parallel \vec v$, the elementary beam becomes narrower
simply because of eq.~(\ref{textbook}).

According to some models of pulsars
(eg.~Melrose 1978; Schopper et al.~2002; Levinson et al.~2005)
the ambient electric field and plasma density within emission region(s) 
in pulsar magnetosphere
tend to oscillate with a frequency $\nuwig$ that can
result either from global magnetospheric electrodynamics (eg.~Sturrock 1971;
Levinson et al.~2005) or can be locally 
excited by streams of energetic electrons
penetrating the ambient plasma (eg.~Schopper et al.~2002).
Hereafter we call this frequency a `wiggler' frequency by a rough analogy
with the laboratory free electron laser (FEL).\footnote{Unlike in the device, 
in the models mentioned above the wiggler field is oscillating parallel
to local $\vec B$-field and electron velocity. Perpendicular
wiggler oscillations have also been considered in the context of pulsars
(see Fung \& Kuijpers 2004).}
In the free electron maser models of coherent radio emission from pulsars
the observed radio emission can be considered as the Compton scattered
(and blueshifted) wiggler frequency $\nuwig$:
\begin{equation}
\nuobs \simeq \gamma^2\nuwig.
\label{nuobs}
\end{equation}
What are the spectra of $\gamma$ and $\nuwig$ is a difficult question,
but one simple and seemingly natural possibility is that the
electron energy distribution is much broader than the Fourier spectrum
of the wiggler oscillations, ie.~$\nuwig\sim const$ in eq.~(\ref{nuobs})
for a given object. 
In this case different $\nuobs$ can be associated with the 
inverse-Compton scattering
of the same wiggler frequency $\nuwig$ by electrons with different
Lorentz factors. This implies (from eqs.~\ref{separation} and \ref{nuobs})
that the separation of double structures 
should decrease with frequency according to 
\begin{equation}
\sep \simeq \sqrt{\frac{4}{5}}\frac{1}{\sin\zeta} 
\left(\frac{\nuwig}{\nuobs}
\right)^{1/2} \propto \nuobs^{-1/2}.
\label{sepnu}
\end{equation}
This dependence is in good agreement with the observed behaviour of the
notches at different frequencies.
Fig.~\ref{merge} presents $\sep$ as a function of $\nuobs$
as derived from the data
on J0437$-$4715
(NMSKB, Jenet et al.~1998) and B1929$+$10 (MR04; this work).
All points except from the two diamonds (J0437 at 1380 MHz)
have $2\sigma$ errors marked.
The locations of the diamonds have been estimated
from fig.~1 in Jenet et al.~(1998). The vertical bars at these points
have the ad-hoc length of $\pm10\%$ of $\sep$.

   \begin{figure*}
   \resizebox{12.7cm}{!}{\includegraphics[width=0.8\textwidth]{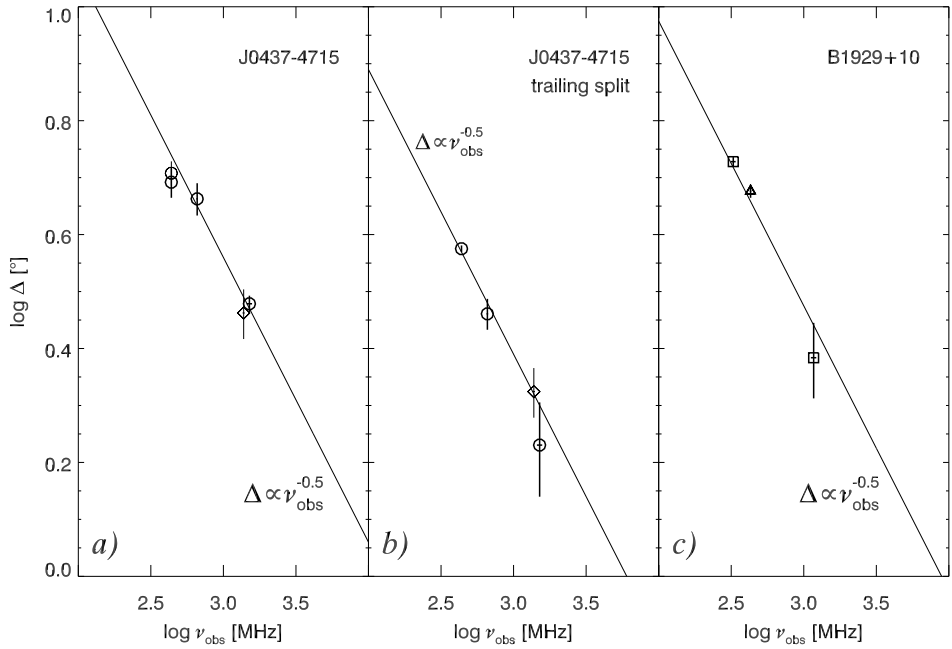}}
   \hfill
   \parbox[b]{50mm}{
      \caption{Observed separation $\sep$ of double notches as a function
      of frequency $\nuobs$ for J0437$-$4715 ({\it a}) and
      B1929$+$10 ({\it c}). The middle panel is for the trailing bifurcated 
      component in J0437$-$4715.
      The error bars are $2\sigma$ for all points except from the two diamonds
      (J0437$-$4715 at 1380 MHz) which are equipped with bars of length 
      $\pm10\%$ of $\sep$.
      The straight lines mark the relation $\sep \propto \nuobs^{-1/2}$
      predicted by eq.~(\ref{sepnu}).
      The light circle for the lowest frequency point of J0437$-$4715
      refers to the separation of notches in the averaged profile
      of linearly polarized radiation L. 
      The data are from NMSKB (circles), Jenet et al.~(1998) (diamonds),
      MR04 (triangle) and this work (squares).
              }
         \label{merge}
	 }
   \end{figure*}

\item[5.] 
The model expressed by eq.~(\ref{sepnu}) assumes that different radio
frequencies are generated by electrons that have different energy
but occupy the same emission region. This is in clear contrast to
the traditional view according to which different $\nuobs$ originate
from different altitudes with different electron plasma frequency, ie.~are
associated with variations of $\nuwig$ in eq.~(\ref{sepnu}). 
Our model predicts then that
locations of emission components in averaged pulse profiles should
not change with $\nuobs$. This is in perfect agreement with 
the frequency independence of the separation between the interpulse (IP)
and main pulse in B0950$+$08 (Hankins \& Cordes 1981). 
The interpulse is connected
with a bridge of low intensity emission with leading components
of the MP. All the components likely have the same origin because
they exhibit similar long term variability (cf.~figs.~1 and 2 
in MR04) as well as similar (low) intensities of single pulses
(Nowakowski 2003).
Therefore, it is probable
that the entire stretch of emission that includes
the IP and the `notched bump' ahead of MP is generated by
the inverse Compton parallel acceleration maser.
\emph{Thus, we are surprised to realize that the issue of whether
the MP-IP separation does (or does not) depend on $\nuobs$ has nothing to do 
with the problem of whether we see one pole or two poles.}
However, since some parallel acceleration models have inherently built-in
two-directional emission (eg.~Levinson et al.~2005), it is still reasonable
to interpret the structure formed by the IP, bridge and MP
in B0950 within the two-directional emitter scenario
(Dyks et al.~2005b; Fowler \& Wright 1982; Cheng \& Ruderman 1977).
This speculation needs to be verified by detailed modelling.

\item[6.]
For the same reason (broad band radio emission caused by electron 
energy distribution)
the conal-like components in J0437$-$4715
do not exhibit any sign of radius-to-frequency mapping over a very
wide frequency range (McConnell et al.~1996; NMSKB).
McConnell et al.~(1996) report an intrinsic low-frequency spectral
turnover around $\sim 100$ MHz which can be associated within our model
with a lower boundary of the electron energy distribution at 
$\gamma_{\rm min} \simeq 6/\sin\zeta$. It is worth mentioning that
below $\nuobs \simeq \nuwig\gamma_{\rm min}^2$ the relation
$\sep \propto \nuobs^{-1/2}$ is not expected to hold.

\item[7.] The oscillations of ambient $\vec E$  may be far from stable 
on long time scales. Evolution or disturbances 
of the oscillation frequency
(broadening of the oscillation spectrum)
would smear the notches.  
This is qualitatively consistent 
with the observed
temporal variations of the shape of double notches (MR04; 
did Phillips 1990 see the notches? -- see fig.~2 therein,
and comments in Rankin \& Rathnasree 1997 or MR04).

\item[8.] 
If the emission is indeed caused by the parallel acceleration, 
there should be a chance
(provided that the macrosopic geometry of the emission region is suitable)
to see the elementary hollow cone \emph{in emission}. There is indeed
\emph{a bifurcated emission component seen in the same interval 
of emission that
contains the double notches} in J0437$-$4715
(see fig.~1 in NMSKB).
Bifurcated components observed in some pulsars
(eg., also for the millisecond PSR J1012$+$5307, see fig.~1 in
Xilouris et al. 1998, or for the radio magnetar XTE J1810$-$197, 
see fig.~1 in Camilo et al.~2006)
may result from the cut of our line of sight
through emission from such a region.
In Fig.~\ref{emi} we present a modelled lightcurve for an elongated
emission region that was thin in the azimuth direction
($\Delta\phi=0.3\theb$) but very elongated in colatitude
($\Delta\theta=10\theb$). 
This configuration can be considered to be a zeroth order approximation of 
a trailing side of a ring centered at the dipole axis.
One can see that the large meridional extent
of the emission region does not blur the hollow cone shape
of the elementary beam completely. The central panel of Fig.~\ref{merge}
shows that the separation between the maxima of the bifurcated component
in J0437$-$4715 also follow the relation $\sep\propto\nuobs^{-1/2}$.

   \begin{figure}
      \centering
      \includegraphics[width=0.5\textwidth]{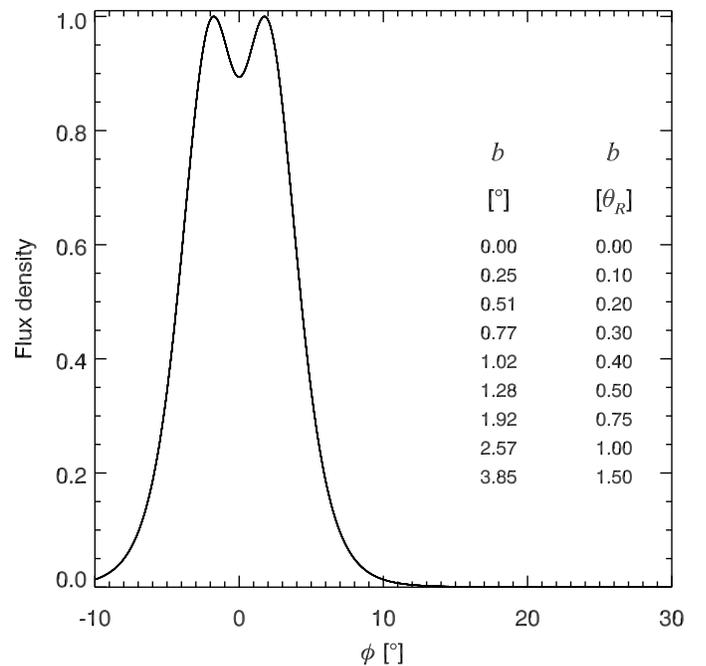}
      \caption{Modelled pulse profile for an emission region considerably 
      extended
      in the rotational colatitude $\theta$ ($\Delta\theta \gg \theb$,
      $\Delta\phi < \theb$).
      The elementary emission beam had a shape
      of the hollow cone for the parallel acceleration case with $\gamma=10$.
      The hollowness is revealed by the $\sim10$\% dip
      at the top of the pulse. The result does not depend on the viewing angle
      $\zeta$.}
      \label{emi}
   \end{figure}

\item[9.] 
The only known radio magnetar -- XTE J1810$-$197 --
exhibits radio emission that is variable on several long time-scales
(days and months, Camilo et al.~2006, 2007) which makes it distinct from 
normal radio pulsars. The shape of its averaged pulse profile 
became extremely unstable in July 2006 (Camilo et al.~2007).
Several months before, however, the object exhibited emission
with many features similar to those of
the `notched emission' of J0437$-$4715.
Its radiation was highly polarized
($89\pm5\%$ at 8.4 GHz, Camilo et al.~2006), the main pulse was bifurcated
and connected smoothly to an extended emission component at low $\nuobs$
(see fig.~1 in Camilo et al.~2006, note that the phase alignment of the profiles
at different $\nuobs$ is arbitrary).
The frequency behaviour of the MP is unlike the normal radius to frequency
mapping: instead of the evolution from a single component at high $\nuobs$
to the well separated two conal components at low $\nuobs$ the MP
becomes broader but retains its basic shape through the very wide
frequency band of $0.7-42$ GHz.
We argue here that the frequency evolution results from the broad 
electron energy spectrum (eq.~\ref{sepnu} with $\nuwig \sim const$).
In Fig.~\ref{magnetar} we show that the separation between the maxima
in the MP approach each other at a rate reasonably consistent
with the $\nuobs^{-1/2}$ law at $\nuobs\ga 1$ GHz. 
The vertical error bars have magnitude of $2\sigma$.\\
At sub-GHz frequencies $\sep$ ceases to increase, probably because
of a `boundary effect' caused by a low energy limit $\gamma_{\rm min}$
in the electron energy distribution. The sub-GHz radiation
may form a low energy end of the radio spectrum with $\nuobs \la 
\nuwig\gamma_{\rm min}^2$. The three flux-density measurements 
between $0.69$ and $2.9$ GHz done on MJD 53850.9 (Table 1 of Camilo et 
al.~2006) are indeed consistent with a harder spectrum
($S_\nu \propto \nuobs^0$) than measured at higher frequencies
($S_\nu\propto \nuobs^{-0.5}$ between $1.4$ and $19$ GHz) on MJD 53857.

\item[10.] Contrary to other models of double notches (Wright 2004; DFSRZ) 
the emission region considered here does not have to be
radially extended (it may extend moderately or not at all).
The large depolarization and extremely complicated position-angle curve
are characteristic features of regions with large radial extension
(Dyks et al.~2004a)
and are not observed in the pedestal emission.
The observed polarization properties of the pedestal emission
thus seem to be more consistent with the present model.
\end{enumerate}

   \begin{figure}
      \centering
      \includegraphics[width=0.5\textwidth]{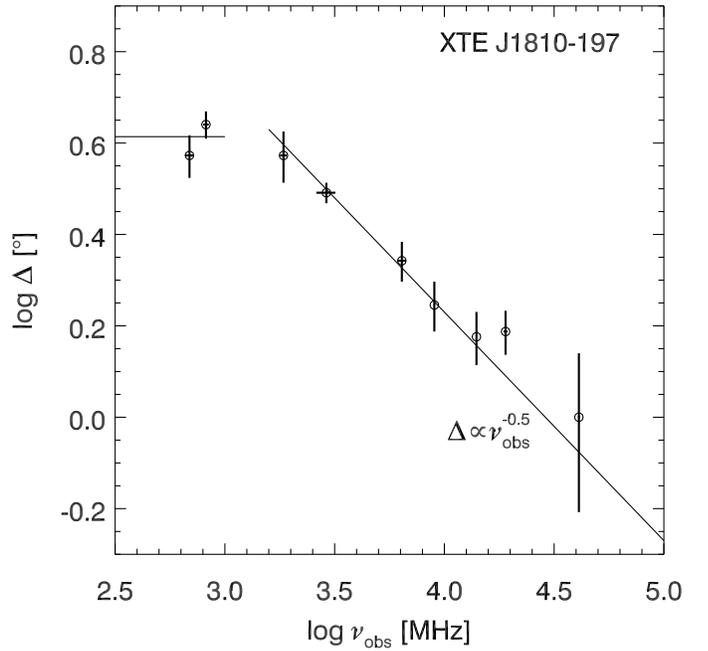}
      \caption{Observed separation $\sep$ of the maxima in the main pulse of
      the first known radio magnetar XTE J1810$-$197 as a function
      of observation frequency $\nuobs$.
      The errors of $\sep$ have magnitude of $2\sigma$. The horizontal bars
      (too short to be visible for most points) show a bandwidth.
      The straight lines mark the relation $\sep \propto \nuobs^{-1/2}$
      given by eq.~(\ref{sepnu}). The data are from Camilo et al.~(2006).
	}
      \label{magnetar}
   \end{figure}
   
\subsection{Predicted periodicity of microstructure} 

The microstructure observed in pulsars exhibits typical timescales
or quasi-periodicities
(eg.,~Cordes et al.~1990; Lange et al.~1998) 
that used to be interpreted within the FEL 
type models of pulsars as a direct result
of the wiggler oscillations and are used
to estimate the value of the Lorentz factor in these models.
Since in our model $\gamma$ can be independently estimated from the
notches' separation, the model can be verified by checking
whether the typical timescale of microstructure is equal to
\begin{equation}
\twig=\frac{1}{\nuwig}
\simeq \frac{\gamma^2}{\nuobs} = \frac{0.8}{\nuobs\ \sep^2\sin^2\zeta}. 
\label{nuwig}
\end{equation}
Had it been possible to observe both the double notches
and the periodic microstructure with the timescale $\twig$,
one could use the above equation to constrain $\zeta$.

The width of the wiggler spectrum allowed by the model can be assessed
as follows: If we assume that the notches get blurred when $1/\gamma$ changes 
``significantly" (say by a factor of 2), then the spectrum of $\nuwig$
cannot extend by a factor larger than 4 
(since $1/\gamma \propto \nuwig^{1/2}$). This rough estimate applies only to the
part of magnetosphere that produces detectable radio emission.

For B1929$+$10 we observe
$\sep\simeq5.3^\circ$ at $327$ MHz, ie.~$\gamma = 9.7/\sin\zeta$ and
$\twig = 0.28\ \rm \mu s\ \sin^{-2}\zeta$ 
which requires single pulse observations
with sub-microsecond time resolution. Given the weakness of the pedestal
emission in B1929$+$10 this observation can be unfeasible even for SKA.
In the other pulsars, however, the emission with notches 
(or with double emission features) is much brighter 
(a few -- $30$ percent of $\imax$) and should be easier to observe.
For B0950$+$08 one gets $\sep \simeq 5.3^\circ$ at $430$ MHz (MR04)
so that $\gamma \simeq 9.7/\sin\zeta$ and $\twig=0.22
\ \rm \mu s\ \sin^{-2}\zeta$.
The maxima of the bifurcated trailing component in the pulse profile of 
J0437$-$4715 are separated by some $3.9^\circ$ at 438 MHz (NMSKB).
This gives $\gamma \simeq 13 / \sin\zeta$ and $\twig \simeq 0.4
\ \rm \mu s\ \sin^{-2}\zeta = 1.6 \ \rm \mu s\ (\sin30^\circ/
\sin\zeta)^{-2}$.
Attempts to reveal the periodicity of microstructure 
in J0437$-$4715 have been done
at two widely separated frequencies (327 MHz, Ables et al.~1997, and
1380 MHz, Jenet et al.~1998) with apparently conflicting results.
The periodicity of $22$ $\mu$s reported by Ables et al.~would be consistent
with eq.~(\ref{nuwig}) for a very small viewing angle 
$\zeta \sim 8^\circ$ that we consider somewhat extreme.
Jenet et al.~(1998) report no microstructure periodicity down to $80$ ns
but their single pulse signal could have been dominated by the bright core
component with little contribution from the `notched emission' that is 
weak at $1380$ MHz.

The ``absorber version" of our model allows for a range of possible
microstructure periodicity, because the counterpart of eq.~(\ref{nuwig})
becomes additionally dependent on geometric parameters of the system
(eg.~the distance of the absorber from the emission region, the curvature
of the region, etc). Whereas the $\nuobs^{-1/2}$ dependence can appear
in this case under some conditions (eg.~in radial $\vec B$-field), 
a specific $\sep$ translates in this scenario
into a larger opening angle $\theb$ and smaller $\gamma$
(see Fig.~\ref{idea}). The expected $\twig$ are therefore smaller
than in the case shown in Fig.~\ref{idea}a.

\subsection{Single pulse visibility of double notches and emission cones}
\label{secsingle}

Our numerical simulations assumed that the radio emissivity
was uniform and steady throughout the entire emission region,
because we were modelling double features in \emph{averaged}
pulse profiles.
A natural question is whether the double notches can be observed
in single pulse emission. A related question is whether
the hollow cone shape of the elementary radio beam can be directly observed
as pairs of emission features in the instantanenous (single pulse)
radio emission.
The simple answer to these questions is `no' --
the notches that are pronounced in an average profile should not be
seen as an absorption feature in the single pulse data. The hollowness
of the emission cone that can be evident in the averaged pulse
profile (Fig.~\ref{emi}) cannot be recognized in single pulses either.
The reason is that
the instantaneous emissivity within the emission region is very 
non-uniform and variable as suggested by features observed 
on a variety of timescales shorter than $P$ 
(eg.~Weltevrede et al.~2006; Edwards et al.~2003; Johnston \& Romani 2002;
Cairns et al.~2004)
and as is normally assumed in the models of drifting subpulses 
(Ruderman \& Sutherland 1975; Wright 2003). 

The detailed discussion is deferred to Appendix \ref{appen2}.
Here we only mention the timescale relevant to the problem
to isolate some limiting cases.
To simplify the analysis we assume the equatorial geometry illustrated 
in Fig.~\ref{datafig} ($\alpha=\zeta=90^\circ$).
Let us consider a single bunch of electrons (a spark)
associated with a single hollow cone of radio emission of opening angle
$2\theb$. Let us initially assume that the bunch is `frozen',
ie.~it does not move in the corotating frame and its emission cone
does not evolve.
If the leading side of its cone is directed
towards the observer at some moment $\til$ then a considerable
period of time 
\begin{equation}
\dtrot \simeq \frac{2\theb}{\Omega} \simeq \frac{P\sep}{2\pi}
\simeq 7\cdot10^{-5}\ \rm s\  \frac{P}{5 \rm ms}\frac{\sep}{5^\circ}
\label{dtrot}
\end{equation}
must pass before the rotation of the magnetosphere directs the trailing side
of the cone towards the observer, at moment $\tit$.
To detect both sides of the cone the fictious frozen bunch would have to
survive for $\dt \ga \dtrot \sim 7\ 10^{-5} - 3\ 10^{-3}$ s, where
the range corresponds to $P\simeq 5-250$ ms observed among the pulsars 
with notches. The corresponding light travel distance is $\drrot =
c\dtrot \sim (2 - 105)\cdot 10^{6}$ cm, ie.~of the order of, or much
larger than the neutron star radius $\rns$.
In Appendix \ref{appen2} we discuss two types of limitations that 
make the observability
of the notches in single pulses improbable/impossible: one is purely 
geometrical and associated with the non-uniform illumination
of the radially thin emission region (slowly drifting bright spots
that do not evolve on timescales shorter than $\dtrot$
nor do they move relativistically in any direction).
The other one is due to special relativistic kinematics and refers
to a realistic situation of fast moving (outflowing) bunches of electrons
that evolve on a timescale $\dt \ll \dtrot$ that is 
too short for a bunch to expose both sides of its elementary emission cone
to the observer.

\section{Conclusions/Discussion}

We conclude that a remarkable number of peculiar observational effects
can be understood within a radio coherency model that is 
based on inverse-Compton scattering of a single `wiggler' frequency
(or a narrow band of wiggler frequencies) 
by a broad energy distribution of electrons. 
The model explains the `W'-like shape of the notches, the bifurcated
emission components, the convergence of these features at the rate
$\nuobs^{-1/2}$, the frequency independence of the separation between the
IP and MP in B0950$+$08 as well as the lack of radius-to-frequency 
mapping in J0437$-$4715. The model performs reasonably well for the only known
radio magnetar XTE J1810$-$197, which implies that the same mechanism
of coherent radio emission is operating in objects with so wildly different
surface magnetic fields as the magnetars and millisecond pulsars.
This finding is consistent with the linear acceleration origin of radio 
emission, because the strength of $\vec B$-field is largely irrelevant
for this mechanism (Rowe 1995).

There are many unsolved puzzles that remain and that need to be addressed
in future. These are 
the depth of the notches, the macroscopic geometry of the emitter
and hole/absorber, the nature of the hole, etc.
They are closely interrelated so it may be worthwhile to address all of them
simultaneously rather than treat them separately. 
The large depth of double notches 
could easily be produced if the coherent radio emission had
occured at two small angles ($\sim$$1/\gamma$) with respect to the plane of 
B-field lines.
When smeared along the B-field lines this would result in a kind of 
two-planar emission geometry. The maser based on curvature drift (Zheleznyakov 
\& Shaposhnikov 1979; Luo \& Melrose 1992, hereafter LM92) 
offers this kind of emission geometry
(see the bottom panel of fig.~1b in LM92) but it predicts
much weaker frequency dependence of $\sep$ than observed
($\sep \propto \nuobs^{-1/3}$ according to eq.~(9) of LM92).
In the case of models based on parallel acceleration, part of the hollow cone beam
near the B-field line plane would have to get absorbed to obtain the two-planar
emission geometry.
An interesting question
is whether the emission is outward or perhaps inward, as suggested by the
leftward shift of the position-angle curve. 
It is worth emphasizing that two-directional emission is inherent in
some of the parallel acceleration models (eg.~Levinson et al.~2005),
although bulk of a magnetosphere of fast rotating pulsars (J0437$-$4715) has been 
estimated opaque for the backward emission (Luo \& Melrose 2006).
An important issue is the altitude of the emission.
The typical radio frequency expected in free electron maser models
is proportional to electron plasma frequency and
is too high in comparison with the observed $\sim100$ MHz emission
(Kunzl et al.~1998; Melrose 2000). This favors large emission distance
($r \sim \rlc$) which is consistent with the far-from-MP location of the pedestal 
emission (provided that B0950$+$08 or B1929$+$10 
have large dipole inclinations, which seems probable).
The strength of the guiding magnetic field required to stabilize 
parallel Langmuir waves is too low ($10^3$ G in Schopper et al.~2002)
to constrain possible emission site.
On the other hand, the
nearly RVM shape of the position-angle curve (Rankin \& Rathnasree 1997)
suggests low emission altitudes and these would require
radiation from the closed field line region.
The region
is a place of copious pair production according to the outer gap
model (Cheng et al.~1986; Hirotani et al.~2003; Takata et al.~2006; 
Wang et al.~1998, see fig.~3 therein), which is quite successful
in reproducing gamma-ray pulse profiles of pulsars (eg.~Romani 
\& Yadigaroglu 1995; Dyks \& Rudak 2003).
The details of the macroscopic geometry of the system remain a puzzle,
and we emphasize that
the configurations shown in Fig.~\ref{idea}
may be very far from reality. 

A big unsolved issue is what is the relation of the radio emission
considered in this paper to the more normal (?) emission that 
can be classified within the scenario of core and conal beams 
that exhibit nulling, drifting as well as the radius to frequency mapping.
Can the RFM-exhibiting conals be interpreted within the same `parallel FEL'
model, but dominated by variations of $\nuwig$ in eq.~(\ref{nuobs})?
What is the origin of core emission?

   \begin{figure}
      \centering
      \includegraphics[width=0.5\textwidth]{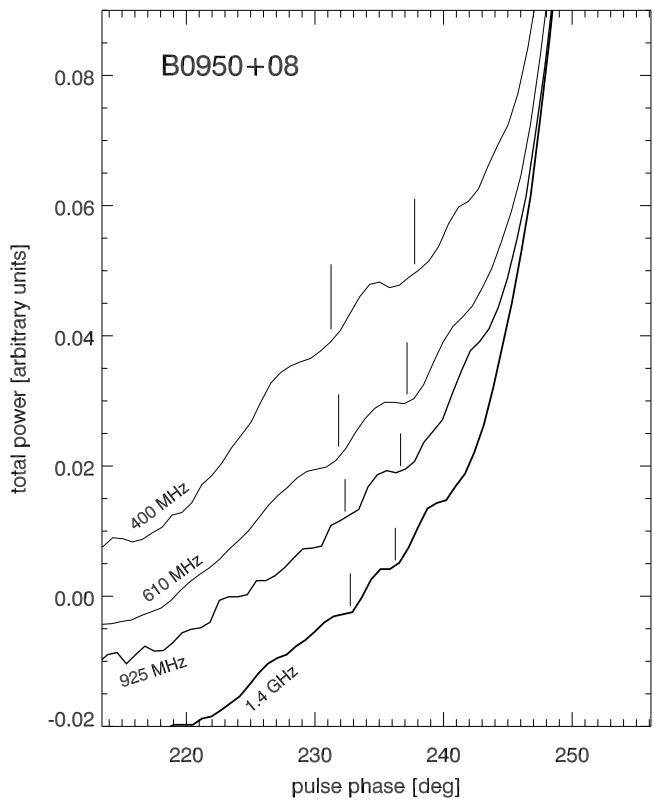}
      \caption{A fragment of pulse profile of B0950$+$08 that shows the 
              behaviour of double notches at increasing 
               frequency (top to bottom). The data are from the European
	       Pulsar Network base (Gould \& Lyne 1998).
	       Pairs of vertical bars above each profile present the relation
	       $\sep = 6.5^\circ(408\ \rm MHz /\nuobs)^{1/2}$
	       (eq.~\ref{sepnu}).
	}
      \label{b09}
   \end{figure}

The model proposed here can be tested observationally by searching
for microstructure periodicities in the emission with notches or
with bifurcated components (eq.~\ref{nuwig}). The expected timescale of the 
microstructure is of the order of $1$ $\mu$s. Multifrequency 
observations of other objects at a high signal-to-noise ratio 
can provide further
support for the relation $\sep \propto \nuobs^{-1/2}$. 
Since there is some evidence of temporal evolution of double notches,
their separation at a fixed frequency may in principle vary in time.
Therefore, \emph{simultaneous} multifrequency
observations would be most preferred, 
although even the non-simultaneous data in Figs.~\ref{merge}
and \ref{magnetar} proved successful in revealing the $\sep(\nuobs)$
relation.
The millisecond pulsar 
J1012$+$5307 has a bifurcated component and is a very good candidate 
(see.~fig.~5 in Kramer et al.~1999). 
The double notches of B0950$+$08 present a difficult observing target
(Fig.~\ref{b09}) and it is important to determine their multifrequency
behaviour with better definition.
The broad band approach can also
tell us whether breaks in the relation
$\sep(\nuobs)$ are associated with spectral breaks. 
A very important but difficult task is to precisely determine
baseline levels for the pulsars with notches at different frequencies.
This would provide the depths of notches that strongly constrain
possible geometric configurations of the magnetospheric emitter.


\begin{acknowledgements}
This paper was inspired by the recent
works of Geoffrey A. E. Wright (MNRAS 2003, 2004). 
JD thanks J.~Borkowski, F.~Camilo, Q.~Luo, and D.~Melrose for helpful comments.
I am also grateful to R. Manchester and S. Ransom for the amazing pulsar data.
We acknowledge the support of the KBN grant N203 017 31/2872
and NSF grants (JMR).
Arecibo Observatory is operated by Cornell
University under contract to the US National Science Foundation.
Part of this research has made use of the data base of published pulse
profiles maintained by the European Pulsar Network, available at:
http://www.mpifr-bonn.mpg.de/pulsar/data/. We have also used the ATNF pulsar
Catalogue (http://www.atnf.csiro.au/research/pulsar/psrcat).
\end{acknowledgements}

\begin{appendix}
\section{The pitch angle case}
\label{pitchang}

The pitch angle case is discussed here for the sake of completeness,
as well as in view of some positive features it has (see below).
We begin with a short description of numerical results
and then discuss how the model performs in confrontation with reality.

   \begin{figure*}
   \resizebox{12.7cm}{!}{\includegraphics[width=0.8\textwidth]{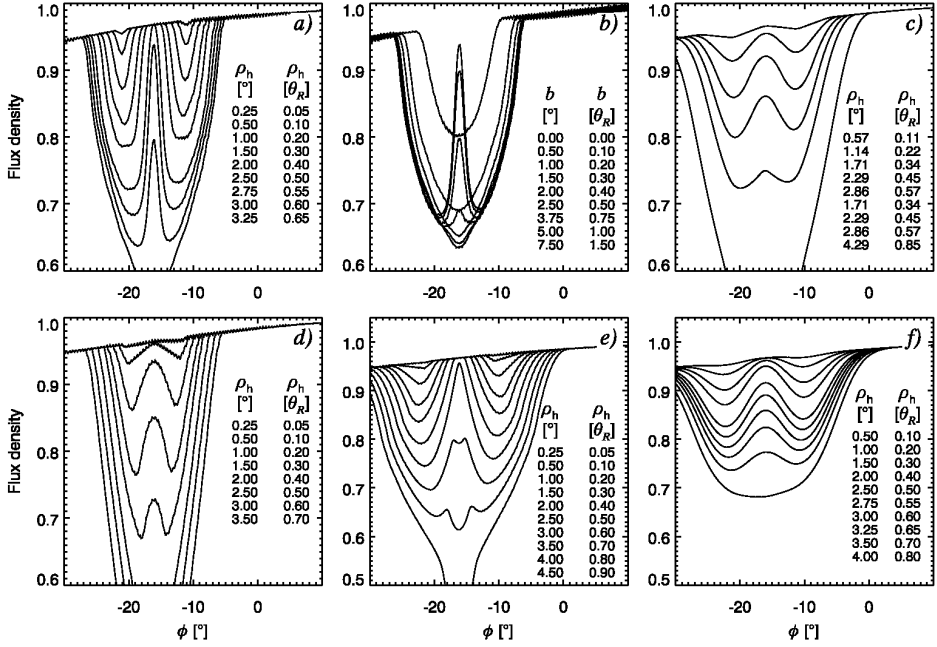}}
   \hfill
   \parbox[b]{50mm}{
      \caption{Modelled double notches calculated for various pitch-angle
      distributions with a fixed $\gamma=10^3$ and $\psimax=5^\circ$. 
      A hole in the emission region is assumed. For details see text.}
         \label{pitches}
	 }
   \end{figure*}

As the simplest choice of the electron's pitch-angle distribution
$\npsi$
we take the triangular shape:
\begin{equation}
\begin{array}{lcrcl}
\npsi = 0  & \ \ \ \  {\rm for} \ \ \ \ &  0 \le & \psi & \le \psimax-\dpsi\\
\lefteqn{\npsi = \cin[\psi - (\psimax-\dpsi)]\dpsi^{-1}\null}  &   & \nonumber\\
& {\rm for} & \psimax-\dpsi \le & \psi & \le \psimax \\
\lefteqn{\npsi = \cout[\psi - (\psimax+\dpsi)](-\dpsi^{-1})}\nonumber\\
  &  {\rm for} & \psimax \le & \psi & \le \psimax+\dpsi\\
\npsi = 0 & {\rm for} & \psimax +\dpsi \le & \psi & \le \pi.
\label{triangle}
\end{array}
\end{equation}
where $\cin$ and $\cout$  can take the values of 0 or 1
and $\dpsi$ is the bottom width of the distribution.
For $\cin = \cout = 1$ the value of $\npsi$ 
increases linearly from $0$ at $\psimax-\dpsi$ up to $1$ at $\psimax$,
then drops linearly down to zero at $\psimax+\dpsi$.
For $\cin = 0$ (and $\cout = 1$) only the outer part of the distribution 
is present with a sharp inner boundary at $\psimax$. For $\cout = 0$
(and $\cin=1$) the inner part of the distribution preserves.

Fig.~\ref{pitches} presents the look of notches for various types
of the pitch-angle distribution $\npsi$. In \ref{pitches}a we have assumed
a `two-sided' distribution ($\cin=\cout=1$) with
$\psimax = 5^\circ$ and $\dpsi = 0.2\psimax$ which is an example of a general 
form of $\npsi$ that is closely confined to $\psimax$ ($\dpsi \ll \psimax$).
This case differs from the parallel acceleration case
mainly in that $\npsi$ covers much smaller solid angle
(in comparison to $\theb^2$) than the beam of non-negligible
emission does in the $\vel \parallel \vec a$ case.
This has two consequences visible in Fig.~\ref{pitches}a:
1) The notches retain the high flux at the W center
for the hole size larger than in the parallel acceleration case
($\rhh \la 0.5\theb$) because no radiation is emitted
near the symmetry axis of the pitch-angle distribution ($\psi < 4^\circ$).
2) The W-shaped notches (with the unaffected central flux) are much
deeper and reach the $\sim 20\%$ depth observed for B1929$+$10.
However, their shape becomes very distorted in comparison 
to the observed one.
For $\rhh < 0.4\theb$ the notches are narrow and separated by a flat plateau 
so that $\sep \gg W$.
For larger $\rhh$ the outer sides of the notches become
rounded and much less steep than the inner ones that form the central 
maximum. All curves in panel a) correspond to the central viewing
of the hole ($b=0$) and $\gamma=10^3$. Fig.~\ref{pitches}b presents result
for the same pitch-angle distribution as in a) but $\rhh$
is fixed to $0.55\theb$ and $b$ is variable.

In Fig.~\ref{pitches}c $\psimax=\dpsi=5^\circ$, ie.~$\npsi$ has a width
comparable to $\psimax$. As before, $\cin=\cout=1$, ie.~the distribution
is symmetrical with respect to $\psimax$. This shape
is qualitatively similar to the shape of the radiation beam
for the parallel acceleration case. Therefore the resulting notches
look alike. The result is for $b=0$.

Fig.~\ref{pitches}d presents results for the inner pitch-angle distribution
($\cout=0$ in eq.~\ref{triangle}) in the case of $b=0$ and variable $\rhh$. 
Here the separation between the notches
is in general different from their width: $\sep > W$ for $\rhh \la 0.2\theb$
whereas $\sep < W$ for $\rhh \ga 0.3\theb$. The notches have peculiar
shape.

Fig.~\ref{pitches}e is for the outer $\npsi$ distribution 
(eq.~\ref{triangle} with $\cin=0$). The shape of notches 
is again unlike the observed one. The W center is flat for 
$\rhh \la 0.5\theb$ or has a dip for $\rhh \simeq 0.75\theb$ and the
notches look strange.

Fig.~\ref{pitches}f is for the pitch-angle distribution
considered by Erber (1973): $\nel \propto \psi\exp\left[
-\psi^2/\psimax^2\right]$. Its general properties 
are similar to the case shown in Fig.~\ref{pitches}c
($\dpsi \sim \psimax$) and therefore the resulting notches
have similar shape, close to the observed one.

The general property that is common for all the cases
is that for very small $\rhh$ the hole acts as a delta function
and the resulting notches have the same shape as $\npsi$.
Therefore, the notches have sharp minima (and are very shallow)
whenever $\rhh \ll \psimax$ (topmost curve(s) in panels a, c, d, and e).
For larger $\rhh$ the minima become oblate in all cases
with the exception of the inner $\npsi$ case (panel d). 
It is worth emphasizing that the \emph{observed} notches 
appear much sharper than they really are on most plots that show
full rotation period (cf.~top line in Fig.~\ref{datafig}a with
exactly the same profile shown in Fig.~\ref{modcomp}a).

A few points on confrontation of the model with the data:

The conclusions 1 to 3 in section \ref{secmodcomp} are valid for both the parallel acceleration case
as well as for a (quite numerous) class of the pitch-angle distributions
that fulfill $\psimax\sim \dpsi$.

The \emph{initial} pitch angle of electron-positron pairs created 
through one-photon
absorption is also inversely proportional to $\gamma$. This is because
photons propagating at small angles $\psi$ relative to $\vec B$
need to have larger energy to produce pairs
[$(\epsilon/mc^2)(B/B_{\rm Q})\sin\psi \approx 2/15$,
Ruderman \& Sutherland 1975]
and the pair components tend to share the energy $\epsilon$
of a parent photon equally ($\gamma mc^2\approx\epsilon/2$, Daugherty
\& Harding 1983), which leads to
\begin{equation}
\gamma \approx \frac{1}{15} \left(\frac{B}{B_{\rm Q}}\right)^{-1}
\frac{1}{\sin\psi}.
\label{gamma}
\end{equation}
Afterwards $\psi$ undergoes strong evolution but its value
remains anticorrelated with $\gamma$ (see fig.~2 in Harding et al.~2005).

The distribution of pitch angles may be considered fragile 
and susceptible to disturbances, which is consistent with
the probable variability of the notches and pedestal emission
on a timescale of days.

The low $|\vec B|$ required for the large $\psimax$ is consistent
with the observed location of the pedestal emission (far from MP
in B1929$+$10 and B0950$+$08) as well as with the presence of 
notches in J0437$-$4715. The magnetar does not seem to fit into the picture,
although we do not know how large the radio emission altitude 
in XTE J1810$-$197 might be.

Since we do not see how the relation $\sep\propto\nuobs^{-1/2}$ could arise in 
the scenario of the favored pitch angle, we consider it less natural
than the $\vec a \parallel \vec \vel$ case.

\section{Single pulse visibility of double features}
\label {appen2}
\subsection{The quasi-steady but nonuniform source at fixed altitude 
(drifting spots)}

As can be seen in Fig.~\ref{idea}b the notches
can be observed in full form in single pulse data
only when the absorber is nearly-simultaneously (or: for a sufficiently
long period of time $> \dtrot$)
illuminated from various directions by a sufficiently large part 
of the emission region (extending at least
between the two points marked with the small letters `d' in Fig.~\ref{idea}b).
In the hole case a nearly-simultaneous emission around the hole is required
(or at least on opposite sides of the hole.)
For a random distribution of sparks (emission spots)
of angular radius $\rho_{\rm sprk} \la \theb$
one can  observe only a part of the notch
feature (a shoulder or dip) or no absorption feature at all. 
The latter should frequently occur because there are a great deal 
of points on the emission region (eg.~marked with `b' in Fig.~\ref{idea})
that contribute unobscured radiation at the pulse phase
of the notches. The phase at which the radiation is detected
is marked also with `b' just above the schematic pulse profiles
at the top of Fig.~\ref{idea}. In this scenario it would be
natural to see no single pulse with the double notches. 
There would also be lots of single emission features 
because of contribution 
of points located outside
of the plane of Fig.~\ref{idea} at angular distance $\sim \theb$
from the line of sight traverse.
One would 
have also seen, however, a great number of double subpulses
corresponding to
the sightline cut through the elementary cones, which is not observed.
The model just discussed is thus excluded by the observed single pulse
properties of pulsar radio emission. In principle, however, such kind of 
emitter could be a part of magnetosphere that stays at a fixed
altitude and is steadily refreshed by electrons that keep flowing 
through it.

\subsection{Bunches of electrons outflowing at relativistic speeds}

A much more realistic model for the basic emission units in pulsar 
magnetosphere is based on short living bunches of electrons
that outflow at relativistic speeds ($\gamma \sim 10$).
Such sources are localized both in the radial and horizontal direction,
but their radial coordinate increases at nearly the speed of light
$dr/dt =0.995c$ for $\gamma=10$. The chance to observe both sides
of the elementary emission cone from a single bunch
fully depends on whether the bunch can survive for sufficiently long
period of time, namely the time interval $\tit - \til$ 
needed for the cone to expose its
other side to the observer.

The source (bunch) is now approaching the observer (and trying to catch up
with the photons) with $\gamma \sim 10$ so that its radial distance
changes according to $r \simeq ct + r_0$. For definiteness, hereafter
we assume that the first side of the cone
is spotted by the observer at time $t_0 =\til = 0$ and at the radial distance
$r_0=r(t_0) \sim \rns$ that will be neglected because $r_0 \ll \drrot$
for most periods we are interested in. Let the azimuth of the emission 
direction on the observed side of the cone is $\phl=0$ (in the observer's 
frame OF), which is also the fixed azimuth of the line of sight.
For a bunch that flows along the dipole axis,
the azimuth of the other side of the cone will change according to
\begin{equation}
\pht \simeq 2\theb - \Omega t - \frac{r}{\rlc} \simeq
\sep - 2\frac{ct}{\rlc} ,
\label{dipax}
\end{equation}
where $2\theb$ is the initial value, $\Omega t$ takes into account
the rotation of the dipole and $r/\rlc$ takes into account the forward
projection of radiation caused by the aberration effect in the limit
$r_0 \ll r$ (eg.~Dyks, Rudak \& Harding 2004b).
The observer can see the trailing side of the cone
($\pht=0$) after time $t=P\sep/(4\pi)$ which is two times smaller
($3 \cdot 10^{-5} - 10^{-3}$ s)
than $\dtrot$ given by eq.~(\ref{dtrot}) and corresponds to light travel 
lengthscales of $(1 - 50)\cdot 10^6$ cm.
The scales are much longer than predicted/considered in some 
models of radio coherency, eg.~$\dr \sim \rm a\ few \times 10^5$ cm in Schopper et al.~(2002)
or in the perpendicular acceleration model of Fung
\& Kuijpers (2004).

Had, however, the bunch survived for the time $\tit - \til$
without any significant evolution (to maintain its original 
hollow cone emission beam) then the trailing side of the cone would \emph{not}
be observed $\sim5^\circ$ after the leading side, but much sooner.
Because the bunch propagates towards the observer with the speed $\vel$ 
close to $c$,
the radiation in the trailing side lags the radio waves from the leading side
by $\dr \simeq c\dt - \vel\dt\simeq c\dt/(2\gamma^2)$. The trailing radio waves
are then detected just after the leading side radio waves:
\begin{equation}
\dtobs = \frac{\tit-\til}{2\gamma^2} = 5\cdot 10^{-8}\ {\rm s}\ \frac{\tit-
\til}
{10^{-5}\ \rm s}\left(\frac{\gamma}{10}\right)^{-2}
\label{lag}
\end{equation}
so that $\Omega\dtobs \ll \sep$.
To detect both sides of the hollow
cone from a single bunch or particle, one would therefore need
to look for double emission features on the timescales of nanoseconds
to milliseconds. The bunch, however, and the physical conditions around it 
could not evolve significantly during the time $\tit-\til\sim 10^{-5}-10^{-3}$ 
s which may be impossible to satisfy. 

We therefore conclude that
the direct observation of the elementary hollow cone in the single pulse data
is extremely improbable, if not impossible. In any case,
in the outflowing bunch scenario the $\sim 5^\circ$ opening angle
of the cone would correspond (in single pulse data)
to double emission spikes separated by $\sim0.01^\circ$. 
The $5^\circ$ separation is visible in the averaged pulse profiles
because the trailing notch 
(or the trailing maxium in Fig.~\ref{emi})
is created by radio waves emitted from roughly the same altitude 
as those associated with the leading notch/maximum, but \emph{emitted later
by a different bunch} of electrons.
Therefore, the only sign of the double notches 
in the single pulse data will be the less frequent appearance
of single-looking, narrow emission spikes that are normally
observed in the high time resolution data.

\subsection{The role of $\vec B$-field line curvature}

The above estimates have been done for the special case of
the dipole axis, but the curvature of magnetic field lines can drag
the other side of the hollow cone into the observer's view.
On the leading side of the dipole axis, the eq.~(\ref{dipax})
becomes:
\begin{equation}
\pht \simeq 2\theb - \Omega t - \frac{r}{\rlc} - \frac{r}{\rhoc}
\label{leading}
\end{equation}
where $\rhoc$ in the last term is the radius of curvature 
of magnetic field lines. The second side of the hollow cone is directed
towards the observer after time
\begin{equation}
ct \simeq \frac{\sep}{\rhoc^{-1} + 2\rlc^{-1}}.
\label{ctleading}
\end{equation}
At the last open field lines $\rhoc \simeq (4/3)(r\rlc)^{1/2}=
9.2\cdot 10^7 (P r_6)^{1/2}$ where $r_6 = r/(10^6\ {\rm cm})$.
For the considered range of periods ($5 - 250$ ms) 
and $r \sim10^6$ cm the resulting
lengthscales are $(0.5 - 5)\cdot 10^6$ cm and become larger
for increasing $r$, eg.~$(0.7 - 12) \cdot 10^6$ cm for $r_6 = 10$.
The lower limit of the scales (that refers to the $5$ ms period) 
is still of the order of $\rns$. The scale has decreased considerably
only for the normal, long period pulsars.

On the trailing side of the dipole axis the sign of the last term in 
eq.~(\ref{leading}) is positive and the equation has no positive
solutions for $t$ if $\rhoc^{-1} > 2\rlc^{-1}$, eg.~near the edge 
of the polar cap. This simply means that the combined effects
of rotation and aberration (2nd and 3rd terms in eq.~\ref{leading})
are too weak to compensate for the backward bending of the $\vec B$-field 
lines, ie.~the field line's curvature drags the beam away from
the observer's line of sight. A full cut through the beam is possible
only when the observer sees the \emph{trailing} side of the cone first.
Appropriate change of signs in (\ref{leading}) gives then
\begin{equation}
ct \simeq \frac{\sep}{\rhoc^{-1} - 2\rlc^{-1}}.
\label{cttrailing}
\label{trailing}
\end{equation}
which gives timescales only slightly larger than on the leading side
because $\rhoc \ll \rlc/2$ in the region where we
make estimates (low altitudes, close to the last open field lines).
An interesting solution of (\ref{cttrailing})
corresponds to $\rhoc = \rlc/2$ (field lines slightly on the trailing side
of the dipole axis) for which $t = \infty$.
This is the case in which the beam does not rotates in the obsever's frame,
ie.~the field line curvature fully compensates aberration and rotation.
The observer can see the leading side of the beam all the time
as the bunch propagates upwards (caustic pile up of radiation
near a fixed phase in pulse profile),
actually up to the altitude above which our simple approximations
(\ref{dipax})-(\ref{cttrailing}) break down.

In the curved magnetic field lines the hollow cone emission caused by
the parallel acceleration becomes dominated by the curvature emission
at frequencies $\nu \la c\gamma^3/(2\pi\rhoc)$ (Melrose 1978).
For the range of $P=5-250$ ms and $\rhoc$ at the polar cap rim
the curvature radiation declines above $0.1 - 0.7$ MHz which is well below the
frequencies at which the notches are observed. 
\end{appendix}


\begin{thebibliography}{}
\bibitem{amd97} Ables, J.G., McConnell, D., Deshpande, A.A., \& Vivekanand, M.
1997, ApJ, 475, L33
\bibitem{bcw91} Blaskiewicz M., Cordes J.M., Wasserman I., 1991, ApJ, 370, 643
\bibitem{cjd04} Cairns, I.H., Johnston, S., \& Das, P. 2004, MNRAS, 353, 270
\bibitem{crh06} Camilo, F., Ransom, S. M., Halpern, J. P., Reynolds, J.,
Helfand, D. J., et al., 2006, Nature, 442, 892
\bibitem{ccr07} Camilo, F., Cognard, I., Ransom, S. M., Halpern, J. P., Reynolds, J., 
et al., 2007, ApJ, submitted (astro-ph/0610685)
\bibitem{cr77} Cheng, A. F., \& Ruderman, M. 1977, ApJ, 216, 865
\bibitem{chr86} Cheng, K. S., Ho, C., \& Ruderman, M. A. 1986, ApJ, 300, 500
\bibitem{cwh90} Cordes, J. M., Weisberg, J.M., \& Hankins, T.H. 1990, AJ 100.2,
1882
\bibitem{dh83} Daugherty J.K., Harding A.K., 1983, ApJ 273, 761
\bibitem{dfs05} Dyks, J., Fr{\c a}ckowiak, M., S{\l}owikowska, A., et al., 2005, ApJ 633, 1101 (DFSRZ)
\bibitem{dzg05} Dyks, J., Zhang, B., \& Gil, J. 2005b, ApJ, 626, L45
\bibitem{dhr04} Dyks J., Harding A.K., \& Rudak B., 2004a, ApJ 606, 1125
\bibitem{drh04} Dyks J., Rudak B., \& Harding, A. K. 2004b, ApJ 607, 939
\bibitem{dr03} Dyks, J., \&  Rudak, B. 2003, ApJ, 598, 1201
\bibitem{esv03} Edwards, R.T., Stappers, B.W., \& van Leeuwen, A.G.J. 2003,
  A\&A, 402, 321
\bibitem{e73} Epstein, R. I., 1973, ApJ, 183, 593
\bibitem{ew01} Everett J.E., Weisberg J.M., 2001, ApJ 553, 341
\bibitem{fw82} Fowler, L.A., \& Wright, G.A.E. 1982, A\&A, 109, 279 
\bibitem{gl98} Gould, D.M., \& Lyne, A.G. 1998, MNRAS, 301, 235
\bibitem{hc81} Hankins, T.H., \& Cordes, J.M. 1981, ApJ, 249, 241
\bibitem{hr06} Hankins, T.H., \& Rankin, J.M., 2006, preprint 
\bibitem{hum05} Harding, A.K., Usov V., Muslimov A.G., 2005, ApJ 622, 531
\bibitem{hhs03} Hirotani, K., Harding, A.K., \& Shibata, S. 2003, ApJ, 591, 334
\bibitem{jak98} Jenet, F.A., Anderson, S.B., Kaspi, V.M., et al. 1998, ApJ,
498, 365
\bibitem{jr02} Johnston, S., \& Romani, R.W. 2002, MNRAS, 332, 109
\bibitem{k70} Komesaroff, M.M. 1970, Nature, 225, 612
\bibitem{kll99} Kramer, M., Lange, Ch., Lorimer, D.R., et al. 1999, 
   ApJ, 526, 957
\bibitem{klj98} Kunzl T., Lesch H., Jessner A., von Hoensbroech, 1998, ApJ
  505, L139
  \bibitem{kis98} Kuzmin, A.D., Izvekova, V.A., Shitov, Yu.P., et al. 1998,
  A\&AS, 127, 355
  \bibitem{lkw98} Lange, Ch., Kramer, M., Wielebinski, R., \& Jessner, A. 1998
  A\&A, 332, 111
  \bibitem{lmj05} Levinson, A., Melrose, D., Judge, A., \& Luo, Q. 2005, ApJ, 631, 456
  \bibitem{lm92} Luo, Q., \& Melrose, D. B. 1992, MNRAS, 258, 616 
  \bibitem{lm06} Luo, Q., \& Melrose, D. B. 2006, MNRAS, 371, 1395 
  \bibitem{lp98} Lyubarski, Y.E., \& Petrova, S.A. 1998, A\&A, 337, 433
  \bibitem{mm01} Malov, I. F., \& Machabeli, G. Z. 2001, ApJ, 554, 587
  \bibitem{mh04} Manchester, R.N., Han, J.L. 2004, ApJ, 609, 354
  \bibitem{mht05} Manchester, R.N., Hobbs, G.B., Teoh, A., \& Hobbs, M.
  2005, Astron. J., 129, 1993
  \bibitem{mab96} McConnell, D., Ables, J.G., Bailes, M., et al. 1996, MNRAS,
  280, 331
  \bibitem{mr04} McLaughlin, M. A., \& Rankin, J. M. 2004, MNRAS, 351, 808 (MR04)
  \bibitem{m78} Melrose, D. B. 1978, ApJ, 225, 557
  \bibitem{m00} Melrose, D. B. 2000, ASP Conference Series, 202, 721
  \bibitem{mns97} Navarro, J., Manchester, R. N., Sandhu, J. S., et al., 1997, ApJ, 486, 1019 (NMSKB)
  \bibitem{n03} Nowakowski, L. 2003, Arecibo Newsletter, 36, 8
  \bibitem{pl85} Perry, T.E., \& Lyne, A. G. 1985, MNRAS, 212, 489 (PL85)
  \bibitem{p02} Petrova, S. A. 2002, MNRAS, 336, 774
  \bibitem{p90} Phillips, J. A. 1990, ApJ, 361, L57
  \bibitem{rc69} Radhakrishnan, V., \& Cooke, D.J. 1969, Astrophys. Lett., 3, 225
  \bibitem{rr97} Rankin, J. M., \& Rathnasree, N. 1997, J. Astrophys. Astron., 18, 91
  \bibitem{ry95} Romani, R.W., \& Yadigaroglu, I.-A. 1995, ApJ, 438, 314
  \bibitem{r95} Rowe, E.T. 1995, A\&A, 296, 275
  \bibitem{rs75} Ruderman M.A., Sutherland P.G. 1975, ApJ 196, 51
  \bibitem{rl79} Rybicki G.P., Lightman A.P., 1979, ``Radiative processes
    in Astrophysics", Wiley-Interscience, New York
    \bibitem{srk02} Schopper R., Ruhl H., Kunzl T.A., Lesch H., 2002, MPE Report
      278, 193
      \bibitem{s71} Sturrock, P.A. 1971, ApJ, 164, 529
      \bibitem{tsh06} Takata, J., Shibata, S., Hirotani, K., \& Chang, H.-K.
       2006, MNRAS, 366, 1310
       \bibitem{wrhz98} Wang, F.Y.-H., Ruderman, M., Halpern, J.P., \& Zhu, T. 1998,
                       ApJ, 438, 314
       \bibitem{wes06} Weltevrede, P., Edwards, R.T., \& Stappers, B.W. 2006,
       A\&A, 445, 243
      \bibitem{w03} Wright, G. A. E., 2003, MNRAS, 344, 1041
      \bibitem{w04} Wright, G. A. E., 2004, MNRAS, 351, 813
      \bibitem{xkj98} Xilouris, K. M., Kramer, M., Jessner, A., et al. 1998, ApJ, 501, 286
      \bibitem{zs79} Zheleznyakov, V. V., \& Shaposhnikov, V.E. 1979, Aust. J. Phys., 32, 49

\end{thebibliography}
\end{document}